# Self-controlled growth of highly uniform Ge/Si hut wires for scalable qubit devices

Fei Gao,[1,2] Jian-Huan Wang,[1,2] Hannes Watzinger,[3] Hao Hu,[4] Marko J. Rančić,[5] Jie-Yin Zhang,[1,2]

Ting Wang,[1,2] Yuan Yao,[1] Gui-Lei Wang,[6] Josip Kukučka,[3] Lada Vukušić,[3] Christoph Kloeffel,[5]

Daniel Loss,[5] Feng Liu,[7] Georgios Katsaros,[3]* Jian-Jun Zhang[1,2]*

1. National Laboratory for Condensed Matter Physics and Institute of Physics, Chinese Academy of Sciences, 100190 Beijing,

China

2. CAS Center for Excellence in Topological Quantum Computation and School of Physics, University of Chinese Academy of

Sciences,100190 Beijing, China

3. Institute of Science and Technology Austria, Am Campus 1, 3400 Klosterneuburg, Austria

4. Frontier Institute of Science and Technology, Xi'an Jiaotong University, 710054 Xi'an, China

5. Department of Physics, University of Basel, Klingelbergstrasse 82, CH-4056 Basel, Switzerland

6. Institute of Microelectronics, Chinese Academy of Sciences, Beijing 100029, China

7. Department of Materials Science and Engineering, University of Utah, Salt Lake City, 84112 UT, USA

Marko J. Rančić is currently at Total@Saclay, Nano-INNOV – Bât .861, 8 Boulevard Thomas Gobert, 91120 Palaiseau, France

Email: jjzhang@iphy.ac.cn, georgios.katsaros@ist.ac.at

Semiconductor nanowires have been playing a crucial role in the development of nanoscale devices for the realization of spin qubits, Majorana fermions, single photon emitters, nanoprocessors, etc. The monolithic growth of site-controlled nanowires is a prerequisite towards the next generation of devices that will require addressability and scalability. Here, combining top-down nanofabrication and bottom-up self-assembly, we report on the growth of Ge wires on pre-patterned Si (001) substrates with controllable position, distance, length and structure. This is achieved by a novel growth process which uses a SiGe strain-relaxation template and can be generalized to other material combinations. Transport measurements show an electrically tunable spin-orbit coupling, with a spin-orbit length similar to that of III-V materials. Also, capacitive coupling between closely spaced wires is observed, which underlines their potential as a host for implementing two qubit gates. The reported results open a path towards scalable qubit devices with Si compatibility.

Semiconductor nanowire (NW) devices have become the workhorse for studying exotic states such as Majorana fermions,[1,2] Andreev bound states,[3,4] for realizing spin qubits,[5,6] programmable NW circuits,[7] nanolasers,[8] and solar cells.[9] Different material combinations ranging from II-VI to IV-IV have been widely used. Among them, Ge is attracting more and more attention due to its high hole mobility,[10-13] low effective mass,[14,15] good contacts with metals,[16-18] strong spin-orbit interactions,[19-22] capability of isotopic purification,[23] and compatibility with Si. These attractive features make Ge a promising candidate not only as a CMOS channel material but also as a host for spin[6,24,25] and even topological qubits.[26,27] Excitingly, the first hole spin qubit[6] and proximity-induced superconductivity with a hard gap[28] have been realized recently in one-dimensional (1D) Ge.

Despite much progress has been made so far, it remains a formidable challenge to have individuals as well as arrays of NWs with a high degree of addressability and scalability for the next generation of NW-based quantum devices. For example, in the field of group III-V semiconductors, precisely positioned NW networks have been achieved with predefined metal islands.[29] The out-of-plane grown NW structures, however, need to be transferred from the growth wafer to a second substrate for device fabrication, which limits their scalability.[7,29] Very recently, impressive in-plane NW networks have been successfully demonstrated with selective-area[30,31] and template-assisted[32] growth techniques, but the problems related to growth imperfections such as dislocation and polytypism still remain. In addition, the selective-area growth works only for material systems which have good growth selectivity between oxides and semiconductors. In the field of group IV SiGe system, attempts have been made on rib-patterned Si (1 1 10) substrate and positioned in-plane Ge wire bundles have been demonstrated.[33] However, isolated wires could not be obtained, which is a prerequisite for scalable quantum devices.

Here, combining bottom-up self-assembly and top-down nanofabrication, we demonstrate the self-controlled growth of highly uniform in-plane Ge wires on Si (001) substrates, which are both addressable and scalable. The so-called Ge hut wires (HWs)[34] grow selectively on an initially formed 1D SiGe layer at the edges of trench-patterned Si. They have a height of about 3.8 nm with a standard deviation of merely 0.11 nm, and their position, distance, length and structure can all be precisely controlled to exhibit an unprecedented high degree of uniformity. Theoretical calculations show that the initially grown 1D SiGe layer provides an enhanced strain relaxation and results in the formation of the Ge HW. Low temperature co-tunneling measurements were performed to determine the spin-orbit coupling (SOC) strength of the Ge HWs. A theoretical model has been developed to extract the SOC length from the experimentally measured singlet-triplet anti-crossing, which concludes that the SOC length of holes in the Ge HWs is comparable to that of electrons in InAs and InSb. Transport measurements further reveal that the SOC length is electrically tunable. In addition, the formation of two closely spaced parallel Ge HWs enables capacitive coupling between quantum dot (QD) devices, which paves the way towards entanglement via two qubit gates.

In order to obtain site-controlled Ge wires, ordered trenches are fabricated by electron-beam lithography and reactive ion etching on an 8-inch Si (001) wafer, as shown by the atomic force microscope (AFM) image in **Figure 1**a. These trenches with an 80 nm width and a 70 nm depth are orientated along either the [100] or the [010] direction (see Methods for a description of the cleaning procedure and the growth details). After the deposition of a 60 nm Si buffer layer and a 3 nm $Si_{0.75}Ge_{0.25}$ alloy at 550 °C, uniform 1D SiGe structures with a trapezoidal cross-section (called from now on mounds) form at the two edges of the trenches as shown in Figure 1b. After the subsequent deposition of 0.6 nm Ge at 550 °C followed by 1h in-situ annealing, Ge wires form on the SiGe mounds (Figure 1c). The initial formation of the 1D SiGe mound is found a prerequisite for the subsequent growth of the Ge HW. The Ge HWs have two (105) facets with an inclination angle of 11.3°, as shown by the surface orientation map in the inset of Figure 1c. This is further confirmed by the cross-sectional scanning transmission electron microscopy (STEM), showing the (105)- faceted cross-section with a height of about 3.8 nm (Figure 1d). Interestingly, as shown in the inset of Figure 1d, the SiGe mound shows an asymmetric trapezoidal geometry (the cross-section is schematically shown in Figure 3a). A sharp interface between the Ge wire and the SiGe mound is observed. No dislocations are observed both in cross-section and along the wire (Figure 1d-e), indicating a perfect single crystal growth of the wires. Higher resolution STEM images are included in the Supporting Information. The height distribution of the Ge HWs is shown in Figure 1f. There is an average height of 3.8 nm and a standard deviation of 0.11 nm. The morphological evolution during the growth is summarized in Figure 1g.

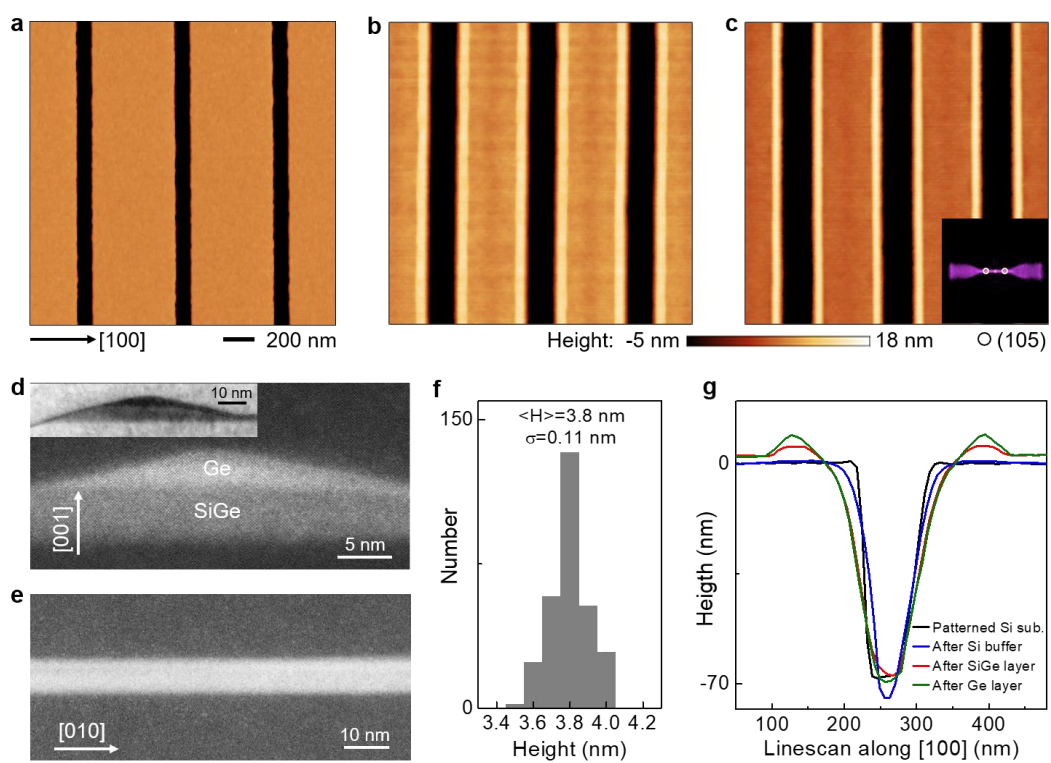

**Figure 1.** Self-assembled growth of site-controlled Ge HWs at the edges of a trench-patterned Si (001) substrate. a-c) AFM images of an [010] orientated trench-patterned Si(001) substrate (a), $Si_{0.75}Ge_{0.25}$ mound structures formed at the two edges of the trenches (b) and Ge HWs grown on the $Si_{0.75}Ge_{0.25}$ mound (c). The black areas in all images are the etched trenches, while the bright 1D structures correspond to the SiGe mounds in (b) and the Ge wires in (c). The inset in (c) shows a surface orientation map demonstrating that the wires are (105) facetted with an inclination angle of 11.3°. d,e) STEM images of a wire in cross-section and along the wire, respectively. The cross-section shows the triangular Ge HW sitting on the SiGe mound. As seen from the inset TEM image in (d) and the red AFM linescan in (g), the SiGe mounds have an asymmetric trapezoidal cross-section. The side next to trench has a slope of 11.3°, while the side away from the trench has a slope of about 9°. f) Histogram showing the height distribution of the Ge HWs. The average height value ⟨H⟩ and standard deviation σ of the distribution are quoted. The Ge HW height is extracted by the measured AFM peak height (green curve in (g)) with the addition of the wetting layer thickness and the subtraction of the $Si_{0.75}Ge_{0.25}$ mound thickness (red curve in (g)). g) AFM linescans along the [100] direction over one trench showing the evolution of the substrate from before growth (black), to after Si buffer (blue) and SiGe layer growth (red) until the final formation of the two site-controlled Ge HWs (green). The 60 nm height offset of the buffer layer has been subtracted from the black trace.

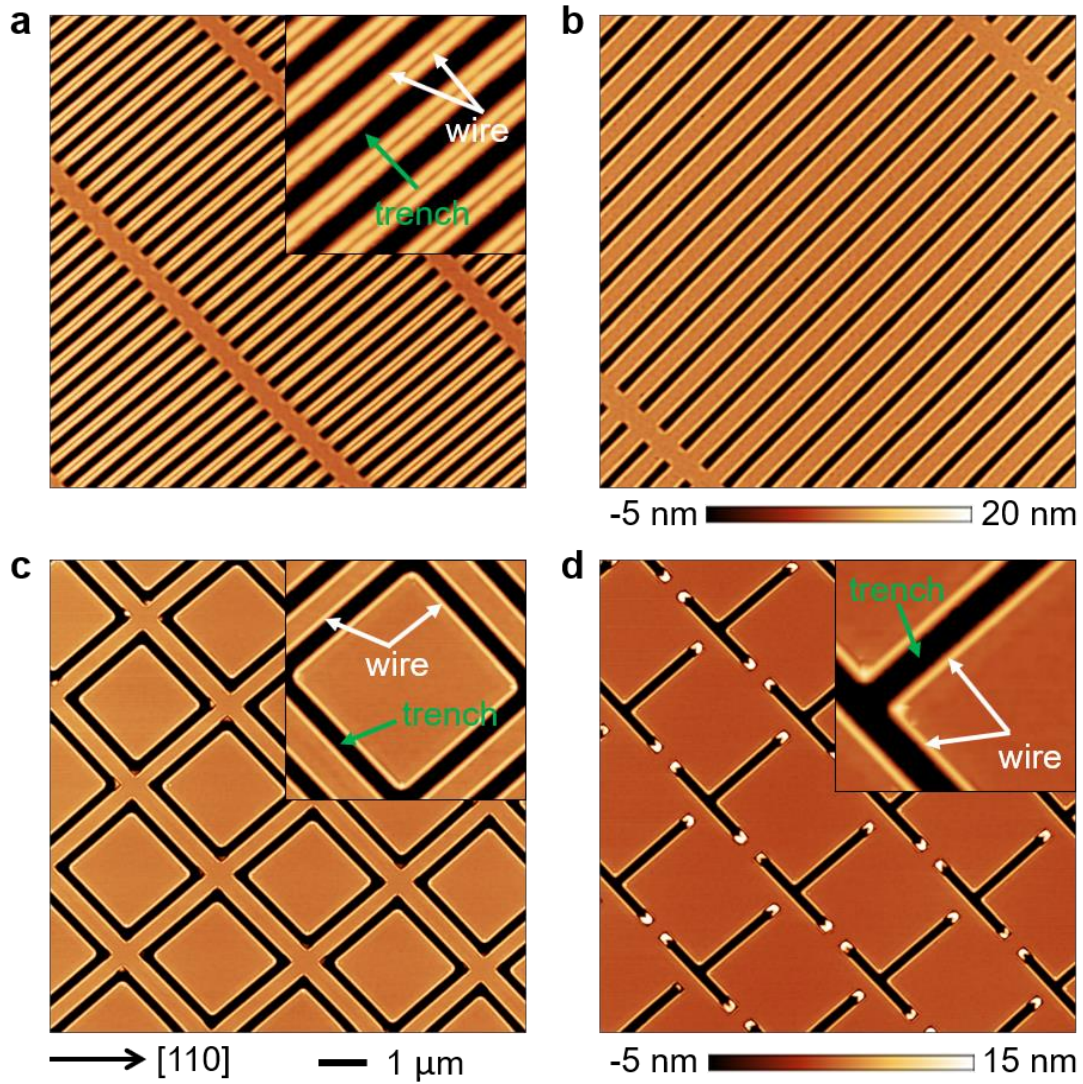

**Figure 2.** Site-controlled Ge HWs with controllable period, length and structure. a) AFM image of parallel HWs grown on a Si substrate with trenches of 4 µm in length and periodicity of 400 nm. The inset shows a zoom-in of the closely-spaced parallel Ge HWs separated by about 30 nm. (b) is similar to (a) but for trenches with a period of 600 nm and a length of 10 µm. c) AFM image of Ge HWs grown on a Si substrate with trenches forming a square shape. The inset shows square-shaped HW structures consisting of four HWs. d) AFM image of Ge HWs grown on a Si substrate with T-like trenches. L-shaped structures consisting of two perpendicular Ge HWs are created. In all images the wires are orientated along the [100] or [010] direction.

As shown above, the wires form at the edges of the trenches, which are fully controllable by the top-down fabrication. We are therefore able to grow HWs with controlled position, distance, length and even structure on a wafer scale. By simply changing the top width of the ridges, the areas between the trenches, we can create two parallel Ge HWs with a neighboring edge-to-edge distance of about 30 nm (**Figure 2**a). This distance is tunable and can be further decreased. However, due to strain repulsion,[35] neighboring wires do not merge together into one wire. The length of the HWs depends only on the length of the trenches, which implies that in principle any length can be obtained. For example, by changing the trench length from 4 µm to 10 µm, we correspondingly obtained ordered Ge HWs with a length of 10 µm (Figure 2b). It is apparent that one can tune the wire period and distance by choosing a different pattern period, as shown in Figure 1a and Figure 1b. Particular geometries like square-shaped structures consisting of four HWs (Figure 2c) and L-shaped structures consisting of two perpendicular HWs (Figure 2d) can also be obtained. For the inner squares (Figure 2c), the wires are connected and form a closed loop, while, for the outer squares, they are mostly disconnected. We expect that connected outer squares can be obtained by tuning the pattern period or the trench sidewall slope. At the ends of the T-shaped trench structure (Figure 2d), one sees larger Ge islands. Their growth is attributed to a larger capture zone of these positions where there are more Ge ad-atoms to diffuse into, resulting in large islands.[36] It could be avoided by decreasing the pattern period. We emphasize that all the HWs are homogeneous with a stable lateral size, not depending on the pattern period, length and structure.

Let us now elaborate on the growth mechanism of the site-controlled Ge wires. When the strained SiGe alloy is deposited on the Si substrate, it first wets the surface of both the flat region and the deep trench. The trench induces notable strain relaxation at the rim[37] which directs the further deposited SiGe to accumulate at the rim to form a SiGe mound. The SiGe mound is seen to have a base size of ~70 nm, which is well below the minimum size required to form the faceted SiGe wire,[34] and hence adopts a shape with continuous changing surface orientation and zero contact angle with the flat region.[38] On the other hand, upon SiGe deposition the upper sidewall of the trench evolves into a shallow (105) facet, which continues into the sidewall of the SiGe mound. Consequently, the SiGe mound adopts an asymmetric shape with the (105) faceted sidewall next to the trench and a shallower sidewall on the other side.

Next, we explain why the Ge HWs form preferably on the SiGe mounds. In general, the growth of strained nanostructures, such as nanoislands, wires or more complex structures, is governed by the competition between surface energy and strain relaxation energy.[39-41] Here we develop a quantitative theoretical model to show that it is the enhanced strain relaxation that drives the Ge HW to grow on a pregrown SiGe mound, rather than on a flat surface (**Figure 3**a,b). We analyze the free energy difference ($dE$) between a Ge HW grown on the SiGe mound and a Ge HW grown on a flat surface. If $dE<0$, the Ge wire is preferred to grow on the SiGe mound; if $dE>0$, on the flat surface.

Figure 3b shows the free energy difference as a function of inclination angle ($\alpha$) and height ($h$) of the SiGe mound (The detailed model and derivation are described in the Supporting Information). One can see that for large enough $\alpha$, $dE$ is negative, meaning that the Ge wire growth on the SiGe mound is energetically favorable; while for relatively small α, $dE$ is positive and the Ge wire prefers to grow on a flat surface. The boundary between the two regions can be determined by letting $dE=0$, which gives the boundary line $h_m$ as a function of α, as shown by the black curve in Figure 3b. The general trend predicted by the model agrees very well with the experiments.

Furthermore, it has been shown that the diffusion barriers of Si and Ge ad-atoms on Si and Ge(001) surfaces increase with increasing compressive strain.[42,43] For pure Ge HW growth (without SiGe mound), the misfit strain of the wetting layer is high, so that the diffusion barrier is larger, favoring the formation of nanoislands. In contrast, by growing the SiGe alloy first, the diffusion barrier is reduced because of a smaller misfit strain, favoring the growth of the very long SiGe mounds. Later when pure Ge is grown, the SiGe mounds act as a "diffusion buffer", so that the diffusion barrier of Ge ad-atom is smaller on the mounds than on the flat surface, which again favors the growth of long Ge HWs.

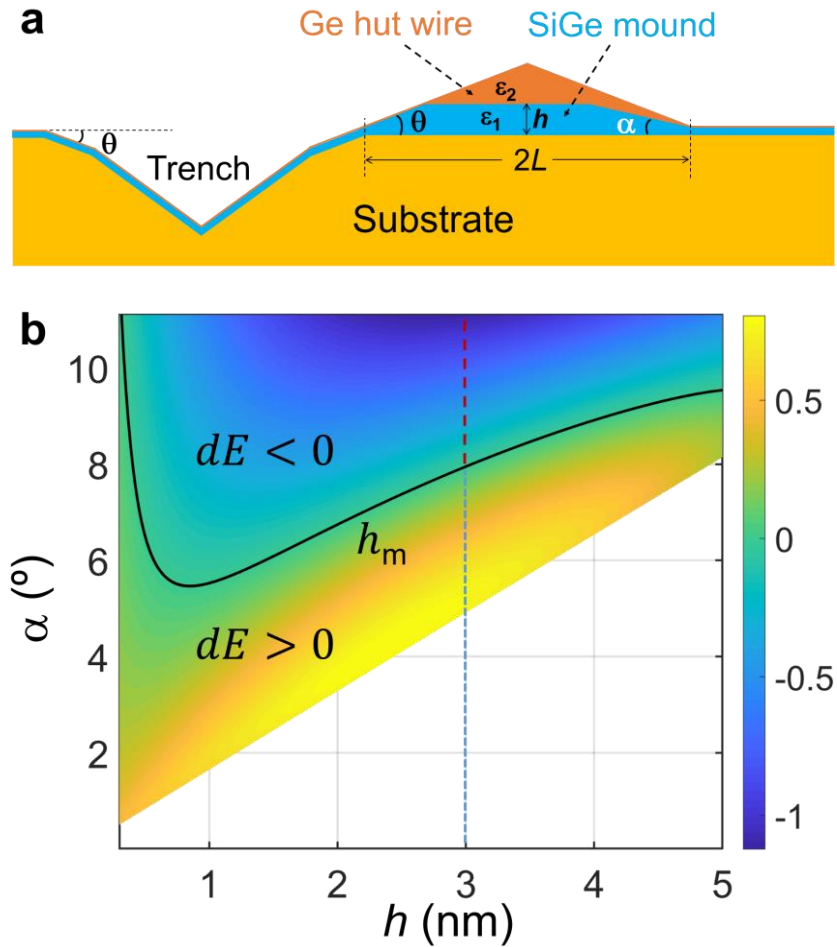

**Figure 3.** Free energy difference between a Ge HW grown on a $Si_{0.75}Ge_{0.25}$ mound and that on a flat Si with the same volume. a) Schematic model for the Ge HW growth on a pregrown SiGe mound. $2L$ is the base size of the pregrown SiGe mound and $h$ is its height. $\alpha$ is the inclination angle of the SiGe mound at the edge away from the trench and $\theta$ is the inclination angle of the SiGe mound at the edge next to the trench and the facet angle of the Ge HW. $\varepsilon_1$ and $\varepsilon_2$ are respectively the mismatch strain of the SiGe mound and the Ge HW with respect to the substrate Si. b) Free energy difference $dE$, in units of eV, vs. mound inclination angle $\alpha$ and height $h$. The black line $h_m$ indicates the boundary between positive and negative $dE$. The dotted line corresponds to the experimental value of the SiGe mound height. $\varepsilon_1/\varepsilon_2=0.25$ is used in the calculation, in accordance with the Ge concentration in the $Si_{0.75}Ge_{0.25}$ alloy.

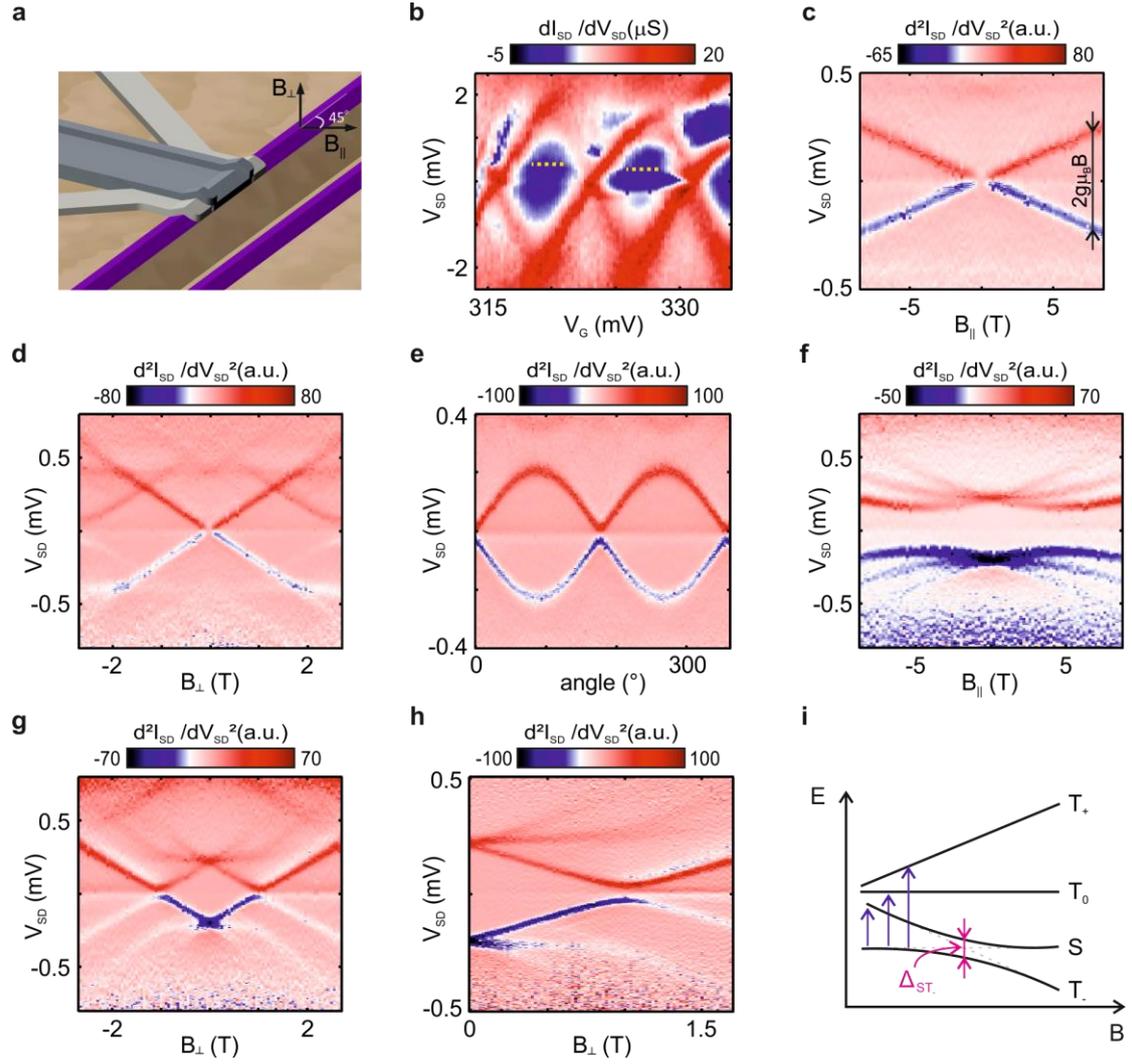

**Figure 4.** Magneto-transport measurements of single QD devices and singlet triplet anti-crossing.

a) Schematic of three terminal device on a site-controlled HW. Light grey electrodes indicate source and drain contacts while the darker grey electrode is the top gate. The insulator is not shown. b) $dI/dV$ versus $V_{SD}$ and $V_G$ at zero magnetic field showing the characteristic diamond plot. Inside the Coulomb diamonds elastic co-tunneling and inelastic co-tunneling steps (marked by yellow dashed lines) can be observed. c,d) $d^2I/dV^2$ versus $V_{SD}$ and parallel (perpendicular) magnetic field $B_\parallel$ ($B_\perp$) for $V_G = 320.9$ mV. The Zeeman splitting of the ground state with an odd hole number can be clearly observed. g-factors of 0.50±0.01 and 3.91±0.02 can be extracted for $B_\parallel$ and $B_\perp$, respectively. e) $d^2I/dV^2$ versus $V_{SD}$ and angle of magnetic field for B=1 T and $V_G = 320.9$ mV. A g-factor anisotropy of about 8, underlying the heavy-hole character of the confined states, [44] is measured. Zero degrees correspond to the parallel magnetic field direction $B_{\parallel}$ and 90 degrees to the out-of-plane field $B_{\perp}$. f,g) $d^2I/dV^2$ versus $V_{SD}$ and $B_{\parallel}$ ($B_{\perp}$) for $V_G$ = 327.9 mV and an even hole QD occupation. h) High resolution measurement similar to (g) for highlighting the singlet triplet anti-crossing at about 1 T. At the anti-crossing the T. state becomes the ground state. i) Energy diagram showing the threefold splitting of the triplet state with magnetic field. Due to spin–orbit coupling, the lowest energy triplet and the spin-singlet state anti-cross.

A key parameter for NWs, both in view of their potential to host Majorana bound states and hole spin qubits is the SOC strength. In order to investigate this important parameter in our system, single hole transistors (**Figure 4**a) were fabricated out of the site-controlled HWs (see methods for details), and low temperature magneto-transport measurements were performed. Differential conductance ($dI/dV$) was measured as a function of the source-drain bias ($V_{SD}$) and the gate voltage ($V_G$). These measurements verified the realization of single hole transistors as closing Coulomb diamonds can be observed in Figure 4b. Elastic co-tunneling and inelastic co-tunneling features (indicated by dashed yellow lines in Figure 4b),[45] leading to finite conductance within the Coulomb diamonds, is present due to the good electrical contacts to the HWs. In inelastic co-tunneling the transition between the ground state and excited states can be observed and thus be used for performing spectroscopy measurements. For a diamond with an odd hole occupation number and for a fixed gate voltage value within the Coulomb blockade regime, the transition between the spin ground and excited state can be seen when sweeping the magnetic field. This allows the measurement of the Zeeman splitting and thus the extraction of the g-factors (Figure 4c and 4d) and their anisotropy (Figure 4e). Furthermore, by changing the number of confined holes by one, an even hole occupancy is achieved. The measurements of $d^2I/dV^2$ as a function of $V_{SD}$ and $B_\parallel$, (Figure 4f), show that the excited state is comprised of three states, with all three splitting non-linearly in the magnetic field. We account this non-linearity in the Zeeman splitting to orbital effects. By changing the direction of the magnetic field, from in-plane ($B_\parallel$) to out-of-plane ($B_\perp$) direction where the g-factor is a factor of 8 larger, a $ST_-$ anti-crossing can be observed around $B_\perp$=1 T (Figure 4g-i). Due to the fact that the g-factor is much larger for the $B_\perp$ direction, we are able to identify the anti-crossing before any orbital effects start to be significant. Such an anti-crossing indicates the presence of SOC in the site-controlled Ge HWs with an anti-crossing ($\Delta_{ST_-}$) of about 35 µeV.

Attention is now turned to a second device for which data at the low and even hole number regime was obtained. We investigate the effect of the applied electric field on the SOC strength. **Figure 5**a-b show the $dI/dV$ vs $V_{SD}$ and $B_\perp$ for increasing gate voltage value. It can be clearly observed that the magnitude of the anti-crossing increases with the applied top gate voltage. We model the electric field dependence of the spin-orbit coupling with the following Hamiltonian

$$H_{SO} = \alpha' k_y \sigma_x.$$

$k_y$ is the momentum (wavenumber) along the HW and $\sigma_x$ is the $x$-component of the spin operator. The partly phenomenological spin-orbit coupling coefficient $\alpha' = (\alpha_{DR} + \alpha)E_z + \beta$ is composed of the part $(\alpha_{DR} + \alpha)E_z$ which depends on the electric field $E_z$ and the part $\beta$ which is independent of the electric field. To elaborate more, $\alpha_{DR}$ is the direct Rashba coefficient which is a function of the dimensions of the wire, material parameters and strain parameters (see Section IV in the Supplementary material); $\alpha$ and $\beta$ are phenomenological parameters that depend, e.g., on the microscopic details of the interfaces.

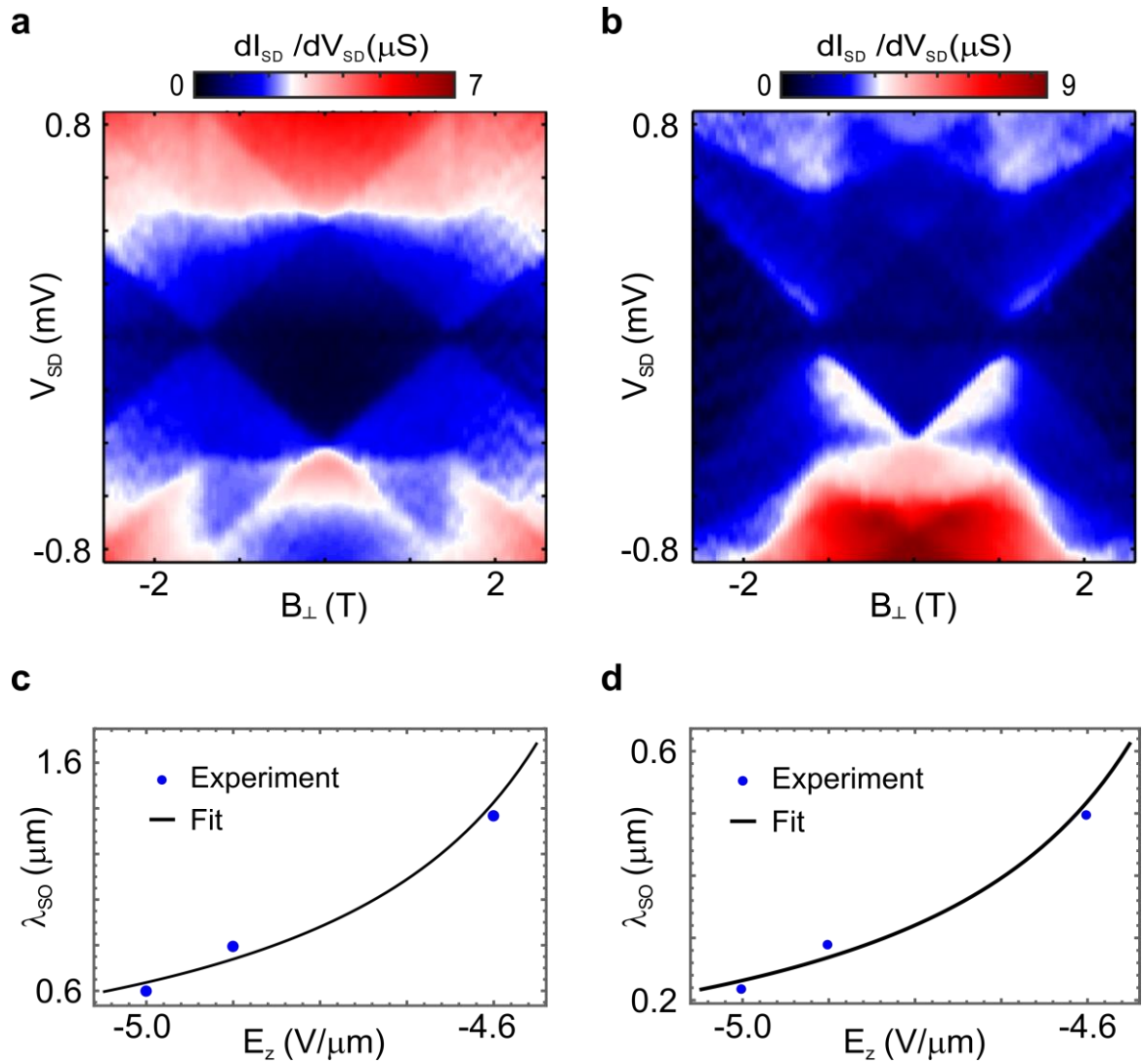

**Figure 5.** Electrical tunability of spin-orbit coupling. a,b) $dI/dV$ versus $V_{SD}$ and $B_\perp$ for $V_G = 510.5$ and 550 mV, respectively. $\Delta_{ST}$ increases from 28 μeV to 60 μeV, occurring at magnetic field strength of 1-2 Tesla. c,d) Spin-orbit length versus the vertical electric field strength for a light-hole (c) and a heavy-hole (d) effective mass.

Our method of calculating the $ST_-$ anti-crossing is based on the approach of [46], adapted for a 1D problem. We start by obtaining an analytical expression for singlet and triplet wave functions of two charged particles in a harmonic oscillator potential, valid in the limit of strong Coulomb repulsion (for more information see Supporting Information material Chapter V and VI). The $\Delta_{ST_-}$ anti-crossing is given by

$$\Delta_{ST_-} = 3^{1/4}\sqrt{2}\frac{|\alpha'|}{\lambda_c},$$

where $\lambda_c$ is the confinement length of the quantum dot. The relation between $\lambda_c$, the orbital level spacing $\hbar\omega$ and the effective mass $m^*$ is $\lambda_c = \sqrt{\hbar/(m^*\omega)}$. Furthermore, the orbital level spacing is related to the $ST_0$ splitting $\Delta_{ST_0}$ through the formula $\Delta_{ST_0} = \hbar\omega\sqrt{3}$ (see Section VI in the Supplementary material, where $\Delta_{ST_0}$ corresponds to $\Delta E_{ST}$). The $ST_0$ splitting $\Delta_{ST_0}$ can be read out from Figure 5a-b, thus allowing us to also obtain the orbital level spacing $\hbar\omega$, while the effective mass $m^*$ is an unknown parameter due to the unknown number of holes confined in the quantum dot. Although the effective mass is an unknown parameter it is reasonable to assume that the effective mass is between the light hole and the heavy hole effective masses in germanium, i.e., $m_{LH} \leq m^* \leq m_{HH}$, with $m_{LH} = 0.042m$ and $m_{HH} = 0.32m$, where $m$ is the free electron mass. After obtaining the value of $\alpha'$ from the magnitude of the anti-crossing $\Delta_{ST_-}$, the SOC length is calculated as $\lambda_{SO} = \hbar^2/(m^*|\alpha'|)$ and these results are displayed in Figure 5c-d. As explained in the Supplementary material, Section VIII, this result is consistent with previous studies.[47,48]

For $m_{LH} \leq m^* \leq m_{HH}$ we obtain $600 \text{ nm} \geq \lambda_{SO} \geq 200 \text{ nm}$ for the maximum

achievable value of the electric field $E_z = -5$ V $\mu m^{-1}$. The spin-orbit length is comparable to values expected in InAs and InSb where the spin-orbit length is $\sim 200$ nm.[48-50] The spin-orbit strength has such a large value although the electric field used in the experiment is not optimized. Further increases in the spin-orbit strength can be engineered by varying the electric field over a larger range of values. Our findings show that $2 \cdot 10^{-11}$ eV $\cdot$ m $\geq (\alpha_{DR} + \alpha)E_z \geq 7.1 \cdot 10^{-12}$ eV $\cdot$ m at $E_z = -5$ V $\mu m^{-1}$ and $-1.7 \cdot 10^{-11}$ eV·m $\leq \beta \leq -6.1 \cdot 10^{-12}$ eV·m indicating an interplay between electric-field-dependent and electric-field-independent SOC mechanisms. For more details on SOC we refer the reader to the Supplementary material Sections IV and VI.

We next focus on the potential of the site-controlled Ge wires for the realization of scalable quantum devices. For this we investigate the capacitive coupling between single QDs formed in parallel wires (Figure 2a), because such a capacitive coupling has been suggested and used for the realization of two-qubit gates in double QDs.[51-53] Devices out of two single QDs facing each other have been fabricated (**Figure 6**a). When sweeping the two gate voltages $V_{G1}$ and $V_{G2}$ of gates G1 and G2 versus each other and measuring the sum of the currents through both QDs in device 1 and 2, a typical stability diagram of a parallel double QD is obtained [see Figure 6b]. At the intersectional points of the Coulomb peaks, shifts can be observed caused by single hole tunneling events in each of the two devices. A zoom-in to the Coulomb oscillations of device 1 (2) is shown in the left (right) panel of Figure 6c. Dashed white lines indicate the positions of the characteristic breaks of the lines. The shifts correspond to about 0.02e and 0.13e for device 1 and 2, respectively, where e is the electron charge. The difference in the shift is due to the different leverarm factor of the two QD gates.[54]

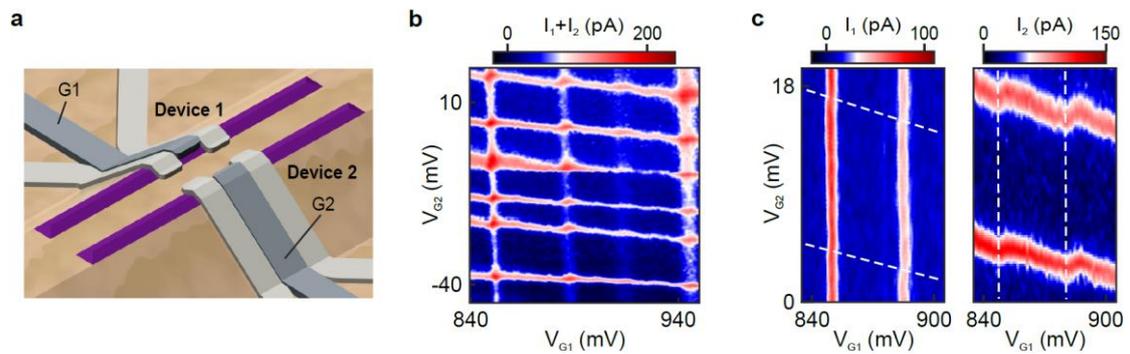

**Figure 6.** Capacitive coupling of site-controlled Ge HWs. a) Schematic showing two three-terminal devices (1 and 2) formed from deterministically grown HWs used for investigating the capacitive coupling between them. b) Total current (I₁+ I₂) versus V_{G1} and V_{G2}. Whenever a single hole tunneling event takes place in one of the devices the Coulomb peaks of the other device shift. c) Zoom-in of (b) showing the current through device 1 (left) and device 2 (right). The white dashed lines indicate the positions of the Coulomb peak shifts.

In summary, in this study we developed a method for monolithic growth of site-controlled Ge wires without the use of any metal catalyst. The method relies merely on strain relaxation via a pre-grown SiGe structure. Such a strain energy relaxation mechanism of a system having a large lattice mismatch via an intermediate layer having a smaller lattice mismatch is general. It is therefore applicable to similar materials including III-V nanowires, which are foremost candidates for topological superconductivity.[1,2] The increase of interest in topologically protected qubits has deemed the necessity of investigating novel Si compatible materials supporting topological superconductivity. While certain studies have focused on Ge in that context,[26] the relatively low in-plane g-factor makes it difficult to reach the topological regime. The solution to this problem could be to use parallel wires as observed here for which it has been predicted that much smaller, if any, magnetic fields will be needed for reaching the topological regime.[27] In addition, the Ge wires covered with Si can suppress metallization,[55] while still allowing proximity induced superconductivity. Finally, the possibility to fabricate wires of arbitrary length, distance and arrangement is crucial for the realization of a recent proposal for Majorana box qubits,[56] where braiding of Majorana qubits is not required for universal quantum control. It therefore becomes clear that our site-controlled Ge HWs on Si are systems where aforementioned proposals can be implemented.

Furthermore, if one envisions scalability one dimensional array of qubits is not the solution, because one non-functional qubit prevents the creation of many-particle entanglement - a requirement crucial for quantum computing. This is why it is of utmost importance for solid state qubits to be in two dimensions, independent of the type of the qubit: spin, charge or topological. The results presented here pave the way for the realization of such scalable arrays of HW qubits in two ways. Directly coupled solid-state qubits can be created within the same HW, while the demonstrated capacitive coupling will allow for generation of long-range entanglement between qubits in parallel wires, leading thus to a two-dimensional array of coupled qubits. In addition, capacitive coupling between parallel wires also allows the creation of charge sensors which are key elements in qubit experiments. Finally, the created closed loops of wires here can be used for implementing Aharonov-Bohm interferometers in solid-state devices and can be thus utilized as magnetometers.[31]

## Experimental section

*Growth*: The trenches are fabricated by electron beam lithography and reactive ion etching on an 8-inch Si (001) wafer. The patterned wafer is cut into $10 \times 10$ $mm^2$ or $16 \times 16$ $mm^2$ small pieces to fit sample adaptors for molecular beam epitaxy (MBE) growth. Before loading into the MBE chamber, the samples are cleaned by the RCA cleaning process and dipped into a 5% hydrogen fluoride (HF) solution for 1 minute to remove the oxide layer and form a hydrogen passivated Si surface. The sample dehydrogenation is first performed at 720 °C in the MBE chamber for 10 minutes. Then the substrate temperature is ramped down to 450 °C for deposition of a 60 nm Si buffer layer with a growth rate of 1 Å s$^{-1}$. The purpose of the homoepitaxial growth of the Si buffer layer is to obtain a high quality surface which was previously damaged by the top-down fabrication. After the buffer layer, a 3 nm $Si_{0.75}Ge_{0.25}$ alloy layer is deposited at 550 °C to form a SiGe mound. The growth rates of Si and Ge are 0.18 Å s$^{-1}$ and 0.06 Å s$^{-1}$, respectively. Keeping the substrate temperature, the Ge growth rate is ramped down to 0.01 Å s$^{-1}$ within 6 minutes. Namely, the $Si_{0.75}Ge_{0.25}$ alloy layer is annealed for 6 minutes and this unintentional process ensures a higher quality of the SiGe mound. Next, a 6 Å Ge layer is deposited at 550 °C followed by 1h in-situ annealing. At last, a 3.5 nm Si capping layer is grown at 330 °C with a rate of 1 Å s$^{-1}$ to protect the Ge HW from oxidation.

*Device Fabrication:* Three terminal devices were fabricated by means of electron beam lithography, metal and atomic layer deposition. As source and drain contacts, 25 nm of Pt were evaporated after an HF dip in order to remove the native oxide. For the gate contacts 3/25 nm of Ti/Pt were deposited on a hafnium oxide layer of 8 nm thickness. In order to demonstrate capacitive coupling of the site controlled Ge wires, we fabricated three-terminal devices from two parallel grown HWs as illustrated in Figure 6a. Both wires are located at the edges of a plateau and are separated by about 30 nm edge to edge.

All the measurements were done with low-noise electronics and in a He-3/He-4 dilution refrigerator at a base temperature of ~40 mK. All lines were filtered at three stages. Pi filters are used at room temperature, LC filters at the mixing chamber stage and a single stage RC filters on the printed circuit board on which the sample was mounted.

## Supporting Information

Supporting Information is available from the Wiley Online Library or from the author.

## Acknowledgements

This work was supported by the National Key R&D Program of China (Grants No. 2016YFA0301701, 2016YFA0300600, and 2015CB932401), the NSFC (Grants No. 11574356, 11434010, and 11404252), the ERC Starting Grant no. 335497, the FWF-Y 715-N30 project and the European Union's Horizon 2020 research and innovation program under Grant Agreement #862046. This research was supported by the Scientific Service Units of IST Austria through resources provided by the MIBA Machine Shop and the nanofabrication facility. F. L. thanks support from DOE (Grant No. DE-FG02-04ER46148). H. H. thanks the Startup Funding from Xi'an Jiaotong University.

## Author contributions

F.G., J.H.W., J.Y.Z., T.W., Y.Y., G.L.W. and J.J.Z. performed the substrate patterning, material growth and characterization experiments. H. W., J. K., L. V. and G.K. fabricated the devices and performed the low temperature transport measurements. H.H. and F.L. developed the growth theory. M. R., C. K. and D. L. performed the transport calculations. J.J.Z. and G. K. planned and designed the experiments.

## Conflict of Interest

The authors declare no conflict of interest.

## Keywords

controllable growth, Germanium, nanowires, qubits, scalability

# Supplementary material for:
# Self-controlled growth of highly uniform Ge/Si hut wires for scalable qubit devices

Fei Gao[1,2], Jian-Huan Wang[1,2], Hannes Watzinger[3], Hao Hu[4], Marko J. Rančić[5], Jie-Yin Zhang[1,2], Ting Wang[1,2], Yuan Yao[1], Gui-Lei Wang[6], Josip Kukučka[3], Lada Vukušić[3], Christoph Kloeffel[5], Daniel Loss[5], Feng Liu[7], Georgios Katsaros[3], and Jian-Jun Zhang[1,2]

[1]National Laboratory for Condensed Matter Physics and Institute of Physics,
Chinese Academy of Sciences, 100190 Beijing, China
[2]CAS Center for Excellence in Topological Quantum Computation and School of Physics and School of Physics,
University of Chinese Academy of Sciences,100190 Beijing, China
[3]Institute of Science and Technology Austria, Am Campus 1, 3400 Klosterneuburg, Austria
[4]Frontier Institute of Science and Technology, Xi'an Jiaotong University, 710054 Xi'an, China
[5]Department of Physics, University of Basel, Klingelbergstr. 82, 4056, Basel, Switzerland
[6]Institute of Microelectronics, Chinese Academy of Sciences, Beijing 100029, China and
[7]Department of Materials Science and Engineering,
University of Utah, Salt Lake City, 84112 UT, USA

## I. STEM IMAGES

A 30 nm Si cap layer was deposited on the hut wires (HWs) to protect the structure during focus ion beam (FIB) cutting. The HAADF STEM was performed using a JOEL ARM200F at 200 kV, with double Cs-correctors.

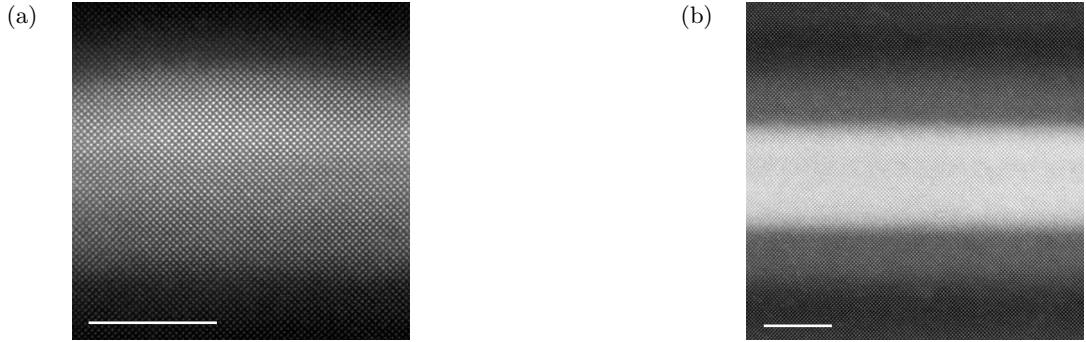

FIG. 1. Filtered high-angle annular dark-field (HAADF) STEM images of the cross-section of the Ge HW (a) and along the Ge HW (b). Scale bar: 5 nm.

## II. GROWTH MODEL

The quantities used in our growth theory are depicted in Fig. 3a in the main text, where $L$ is the half-base-size of the pregrown SiGe mound; $h$ is the height of the SiGe mound; $\alpha$ is the inclination angle of the SiGe mound edge away from the trench, $\theta$ is the angle of the SiGe mound edge next to the trench and the facet angle of the Ge HW; $\varepsilon_1$ and $\varepsilon_2$ are the mismatch strain of the SiGe mound and Ge HW with respect to the Si substrate. We calculate the free energy difference ($dE$) between the free energy of a Ge HW grown on the SiGe mound with that of a wire grown on the flat surface.

The free energy of a Ge wire grown on the SiGe mound can be expressed as $E_m = E - E_1$, where $E$ and $E_1$ are the total energy of a Ge HW grown on the SiGe mound and the total energy of the SiGe mound, respectively; the free energy of the Ge wire grown on the flat surface with the same volume (and the same length) is denoted by $E_0$. Then the free energy difference is $dE = E_m - E_0$.

The total energy of the Ge HW grown on the SiGe mound can be expressed as $E = E_s + E_e$, where $E_s = 2\gamma L / \cos \theta$ is the surface energy, $\gamma$ is the surface energy density of the Ge wire facet. Note that we neglect the Ge/SiGe interface energy for simplicity. $E_e$ is the strain relaxation energy calculated from surface elastic theory [1–3],

$$
\begin{aligned}
E_e &= \frac{1-\nu^2}{\pi Y}\sigma_2{}^2 \Bigg\{ \left(\frac{F_1}{\sigma_2}\right)^2 \left[-\frac{3h_\theta^2}{4} + \frac{h_\theta^2}{2}\ln\left(\frac{h_\theta}{a_0}\right) + h_\theta a_0 - \frac{a_0^2}{4}\right] + \left(\frac{F_3}{\sigma_2}\right)^2 \left[-\frac{3h_\alpha^2}{4} + \frac{h_\alpha^2}{2}\ln\left(\frac{h_\alpha}{a_0}\right) + h_\alpha a_0 - \frac{a_0^2}{4}\right] \Bigg\} \\
&\quad + \frac{1-\nu^2}{\pi Y}\sigma_2{}^2 \left(\frac{F_2}{\sigma_2}\right)^2 \Bigg[-\frac{3(L-h_\theta)^2}{4} - \frac{3(L-h_\alpha)^2}{4} + (L-h_\theta)^2\ln\left(\frac{L-h_\theta}{a_0}\right) + (L-h_\alpha)^2\ln\left(\frac{L-h_\alpha}{a_0}\right) \\
&\quad + \frac{3(L-h_\theta)(L-h_\alpha)}{2} - \frac{(2L-h_\theta-h_\alpha)^2}{2}\ln\left(\frac{2L-h_\theta-h_\alpha}{a_0}\right) + (2L-h_\theta-h_\alpha)a_0 - \frac{a_0^2}{2}\Bigg] \\
&\quad + \frac{1-\nu^2}{\pi Y}\sigma_2{}^2 \left(\frac{F_1 F_2}{\sigma_2^2}\right) \Bigg[\frac{3h_\theta(h_\theta-h_\alpha)}{2} - \frac{h_\theta^2}{2}\ln\left(\frac{h_\theta}{a_0}\right) + L^2\ln\left(\frac{L}{a_0}\right) - (L-h_\theta)^2\ln\left(\frac{L-h_\theta}{a_0}\right) \\
&\quad - \frac{(2L-h_\alpha)^2}{2}\ln\left(\frac{2L-h_\alpha}{a_0}\right) + \frac{(2L-h_\theta-h_\alpha)^2}{2}\ln\left(\frac{2L-h_\theta-h_\alpha}{a_0}\right)\Bigg] \\
&\quad - \frac{1-\nu^2}{\pi Y}\sigma_2{}^2 \left(\frac{F_1 F_3}{\sigma_2^2}\right) \Bigg[-\frac{3h_\theta h_\alpha}{2} + 2L^2\ln\left(\frac{2L}{a_0}\right) + \frac{(2L-h_\theta-h_\alpha)^2}{2}\ln\left(\frac{2L-h_\theta-h_\alpha}{a_0}\right) \\
&\quad - \frac{(2L-h_\theta)^2}{2}\ln\left(\frac{2L-h_\theta}{a_0}\right) - \frac{(2L-h_\alpha)^2}{2}\ln\left(\frac{2L-h_\alpha}{a_0}\right)\Bigg] \\
&\quad - \frac{1-\nu^2}{\pi Y}\sigma_2{}^2 \left(\frac{F_2 F_3}{\sigma_2^2}\right) \Bigg[\frac{3h_\alpha(h_\theta-h_\alpha)}{2} - L^2\ln\left(\frac{L}{a_0}\right) + (L-h_\alpha)^2\ln\left(\frac{L-h_\alpha}{a_0}\right) + \frac{h_\alpha^2}{2}\ln\left(\frac{h_\alpha}{a_0}\right) \\
&\quad + \frac{(2L-h_\theta)^2}{2}\ln\left(\frac{2L-h_\theta}{a_0}\right) - \frac{(2L-h_\theta-h_\alpha)^2}{2}\ln\left(\frac{2L-h_\theta-h_\alpha}{a_0}\right)\Bigg]
\end{aligned}
\tag{1}
$$

where $F_1 = \sigma_1\tan\theta$, $F_2 = \sigma_2\tan\theta$, and $F_3 = \sigma_2\tan\theta - (\sigma_2-\sigma_1)\tan\alpha$; $\sigma_1$ and $\sigma_2$ are the bulk stress in the SiGe mound and Ge HW, both proportional to the mismatch strain $\varepsilon_1$ and $\varepsilon_2$, respectively; $h_\theta = h/\tan\theta$, and $h_\alpha = h/\tan\alpha$; $a_0$ is the cutoff radius, of the order of the interatomic distance. For simplicity, we assume that SiGe and Ge have the same elastic modulus $Y$ and Poisson ratio $\nu$. Then we have $\sigma_1/\sigma_2 = \varepsilon_1/\varepsilon_2$.

The total energy of the SiGe mound is $E_1 = E_{1s} + E_{1e}$, where $E_{1s} = \gamma(2L - h_\theta - h_\alpha + h/\sin\theta + h/\sin\alpha)$. We assume the same surface energy density $\gamma$ for all surface orientations, neglecting the surface energy anisotropy, which will not affect the general conclusion obtained below. The strain relaxation energy $E_{1e}$ of the SiGe mound can be expressed as

$$
\begin{aligned}
E_{1e} &= \frac{1-\nu^2}{\pi Y}\sigma_2{}^2 \Bigg\{ \left(\frac{F_1}{\sigma_2}\right)^2 \left[-\frac{3h_\theta^2}{4} + \frac{h_\theta^2}{2}\ln\left(\frac{h_\theta}{a_0}\right) + h_\theta a_0 - \frac{a_0^2}{4}\right] + \left(\frac{F_4}{\sigma_2}\right)^2 \left[-\frac{3h_\alpha^2}{4} + \frac{h_\alpha^2}{2}\ln\left(\frac{h_\alpha}{a_0}\right) + h_\alpha a_0 - \frac{a_0^2}{4}\right] \Bigg\} \\
&\quad - \frac{1-\nu^2}{\pi Y}\sigma_2{}^2 \left(\frac{F_1 F_4}{\sigma_2^2}\right) \Bigg[-\frac{3h_\theta h_\alpha}{2} + 2L^2\ln\left(\frac{2L}{a_0}\right) + \frac{(2L-h_\theta-h_\alpha)^2}{2}\ln\left(\frac{2L-h_\theta-h_\alpha}{a_0}\right) \\
&\quad - \frac{(2L-h_\theta)^2}{2}\ln\left(\frac{2L-h_\theta}{a_0}\right) - \frac{(2L-h_\alpha)^2}{2}\ln\left(\frac{2L-h_\alpha}{a_0}\right)\Bigg]
\end{aligned}
\tag{2}
$$

where $F_4 = \sigma_1\tan\alpha$.

The free energy of a Ge HW grown on the flat surface is $E_0 = E_{0s} + E_{0e}$, where $E_{0s} = 2\gamma[\sqrt{S/\tan\theta}/\cos\theta - \sqrt{S/\tan\theta}]$; $S$ is the cross-section area of the Ge HW, same as that of the Ge HW grown on the SiGe mound, $S = L^2\tan\theta - h(2L - h_\theta/2 - h_\alpha/2)$. $E_{0e}$ can be expressed as

$$
E_{0e} = -2\ln 2\,\frac{1-\nu^2}{\pi Y}\sigma_2{}^2\tan(\theta)S.
\tag{3}
$$

In the numerical calculation, we treat $L$, $\theta$ and $\varepsilon_2$ as constants, using the experimental value of $L = 35$ nm, $\theta = 11.3°$ for the Ge (105) facet, and $\varepsilon_2 = 0.042$ as the bulk strain of Ge grown on a Si substrate. We use the first-principles calculated value of $\gamma = 6.0$ eV/nm$^2$ for the surface energy density and the elastic energy density is $(1-\nu^2)\sigma_2{}^2/\pi Y = 0.27$ eV/nm$^3$ [4]. $a_0 = 0.5$ nm is used in the calculation, and note that a change of $a_0$ only slightly affects the calculated free energy and will not change the general behavior. The free energy is expressed in energy per unit length (in units of nm) of the wire.

To determine the driving force for the preferred growth of the Ge HW on the SiGe mound, we also analyze the surface energy difference $dE_s = E_{ms} - E_{0s}$ and strain relaxation energy difference $dE_e = E_{me} - E_{0e}$ as functions of $\alpha$ and $h$. As we can see from Fig. 2 (a) and (b), the preferred growth of the Ge HW on the SiGe mound is due to the enhanced strain relaxation, because a mound with large $\alpha$ makes the wire grown on it to have a larger elastic interaction between the Ge wire and the SiGe mound edges.

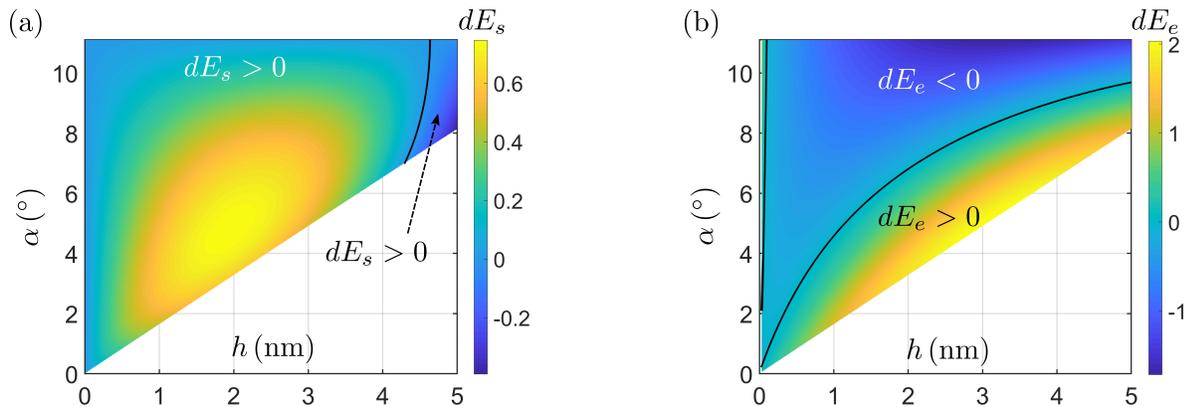

FIG. 2. (a) Surface energy difference $dE_s$ and (b) strain relaxation energy $dE_e$ as a function of the mound inclination angle $\alpha$ and height $h$. The black lines are the boundaries for positive and negative difference regions.

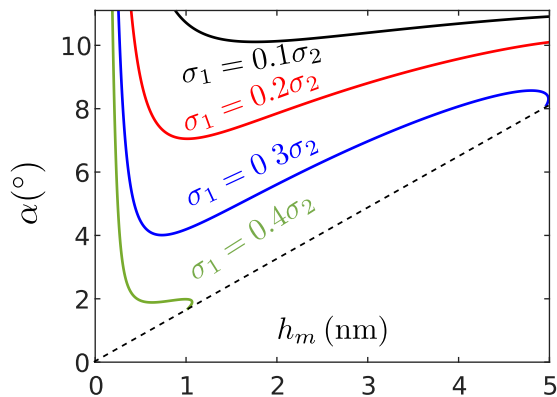

FIG. 3. Plot showing the boundary line $h_m$ as a function of the inclination angle of the SiGe mound $\alpha$ for different values of $\sigma_1$. The dashed line indicates the border to the geometric forbidden region as for given $h$, $\alpha$ can only vary from $arctan(h/L)$ to $\theta$.

As it is shown before, the free energy difference is also a function of the strain/stress in the SiGe with respect to the Si substrate. Fig. 3 shows how $h_m$ varies as a function of $\alpha$ for different values of $\sigma_1$. We can see that the larger the stress in the SiGe mound, the easier the Ge HW grows on the SiGe mound. However, in the experiments, it is not optimal to have the stress in SiGe as large as possible, because if the stress in SiGe is too large, the pregrown SiGe layer may not form at all.

In summary, we demonstrated that the Ge HW growth on the SiGe mound is thermodynamically favorable over the growth on the flat surface. The misfit strain and mound geometry dictate such preference: a larger misfit strain and/or inclination angle will favor the growth on the mound. Another important physical implication of the model is that during heteroepitaxial growth of a Ge wire (or nanoisland) on a Si substrate, the intermixing of Si and Ge at the early stage of growth can significantly affect the growth kinetics by decreasing the formation energy of the nanoisland or wire [5] and hence reducing their nucleation barrier and critical size via the formation of a SiGe alloy core at the base.

## III.   MODELING HOLES IN Ge HUT WIRES

The Luttinger-Kohn Hamiltonian [6]

$$H_{\mathrm{LK}} = \frac{\hbar^2}{2m}\left[\left(\gamma_1 + \frac{5\gamma_2}{2}\right)k^2 - 2\gamma_2\left(k_x^2 J_x^2 + k_y^2 J_y^2 + k_z^2 J_z^2\right) - 4\gamma_3\left(\{k_x, k_y\}\{J_x, J_y\} + \text{c.p.}\right)\right] \tag{4}$$

gives a good description of states at the top of the valence band in Ge, Si, InAs, InSb, and many other materials. Note that Zeeman terms have been omitted in Eq. (4). Here $\gamma_1$, $\gamma_2$, $\gamma_3$ are the Luttinger parameters, $k_x$, $k_y$, $k_z$ are the components of the momentum operator $\boldsymbol{k}$, and $J_x$, $J_y$, $J_z$ are the components of the spin operator $\boldsymbol{J}$. $m$ stands for the free electron mass, c.p. stands for cyclic permutations, and curly brackets denote anticommutators $\{A, B\} = (AB + BA)/2$. The axes $x$, $y$, $z$ coincide with the main crystallographic axes. We choose the $y$ axis parallel to the hut wire and $z$ as the axis of strongest confinement.

Germanium is well described within the spherical approximation $\gamma_2 = \gamma_3 = \gamma_s$, giving rise to the Hamiltonian

$$H_{\mathrm{LK}}^{\mathrm{Ge}} = \frac{\hbar^2}{2m}\left[\left(\gamma_1 + \frac{5}{2}\gamma_s\right)k^2 - 2\gamma_s\left(\boldsymbol{k}\cdot\boldsymbol{J}\right)^2\right]. \tag{5}$$

It is important to note that Eq. (4) can be written in the compact form of Eq. (5) only if the $k_x$, $k_y$, $k_z$ commute, i.e., if $k_i k_j = k_j k_i$. However, in the presence of a magnetic field $\boldsymbol{B} = \nabla \times \boldsymbol{A}$, one finds $\boldsymbol{k} \times \boldsymbol{k} = -ie\boldsymbol{B}/\hbar$ because of $\hbar\boldsymbol{k} = -i\hbar\nabla + e\boldsymbol{A}$, where $\nabla$ is the nabla operator and $\boldsymbol{A}$ the vector potential [7]. Consequently, the form of Eq. (4) must be used for calculations which include corrections due to magnetic-field-induced orbital effects. In our estimate of the effective spin-orbit coupling of holes in hut wires, these corrections are negligible, and so we can set $\boldsymbol{k} = -i\nabla$ and work with Eq. (5) in this Supplementary material.

The influence of strain on the eigenspectrum is calculated by the Bir-Pikus Hamiltonian

$$H_{\mathrm{BP}} = b\left[\epsilon_{xx}(J_x^2 + J_y^2) + \epsilon_{zz}J_z^2\right] = \frac{15}{4}b\epsilon_{xx} + b\left(\epsilon_{zz} - \epsilon_{xx}\right)J_z^2, \tag{6}$$

assuming $\epsilon_{xy} = \epsilon_{yx} = \epsilon_{xz} = \epsilon_{zx} = \epsilon_{yz} = \epsilon_{zy} = 0$ (no shear strain) and $\epsilon_{xx} = \epsilon_{yy}$. Spin-independent terms in Eq. (6) do not affect our results, which is why the hydrostatic deformation potential was omitted. The choice $\epsilon_{xx} = \epsilon_{yy}$ is justified by our COMSOL simulation of strain parameters for a hut wire whose cross section is close to an obtuse isosceles triangle (see Fig. 4). As briefly mentioned above and as depicted in Fig. 4, we will use the following coordinate system throughout this Supplementary material: $z$ is the axis of strongest confinement, $y$ is chosen along the hut wire, and $x$ is perpendicular to these two axes.

For the numerical simulation we considered a hut wire sitting on a SiGe mound along a trench with dimensions as described in the main text. Along the in-plane directions $x$ and $y$ the hut wire is fully strained with $\epsilon_{xx} \approx \epsilon_{yy} = -0.04$ (see the left and the middle panel of Fig. 4). In contrast, the strain tensor element $\epsilon_{zz}$ along $z$ was found to be 0.02 in the center of the wire (see the right panel of Fig. 4).

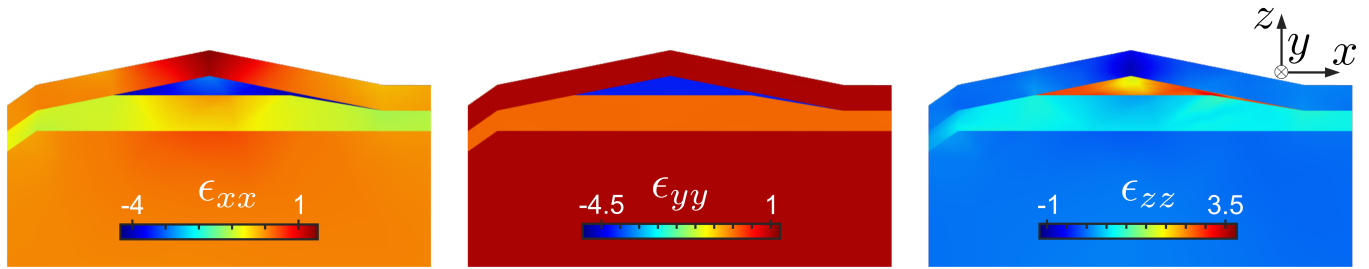

FIG. 4. Elements of the strain tensor from our COMSOL simulation. The strain values are given in %.

As it is hard to analytically obtain the hole spectrum of a hut wire whose cross section is close to an obtuse isosceles triangle, we are going to study a nanowire of rectangular cross section, considering that one side of the rectangle is longer than the other [8]. Adopting the approach of Refs. [8–11], the nanowires are modeled with a hard-wall confinement potential

$$V = V(x, z) = \begin{cases} 0, & |x| < \frac{L_x}{2} \text{ and } |z| < \frac{L_z}{2}, \\ \infty, & \text{otherwise.} \end{cases} \tag{7}$$

Here $L_x$ and $L_z$ represent the length of the sides of the rectangle along the $x$ and the $z$ direction, respectively. The orbital part of a wave function in the $x$-$z$ plane can therefore be written as a linear combination of the basis functions

$$f_{n_x,n_z}(x,z) = \frac{2\sin\left[n_x\pi\left(\frac{x}{L_x}+\frac{1}{2}\right)\right]\sin\left[n_z\pi\left(\frac{z}{L_z}+\frac{1}{2}\right)\right]}{\sqrt{L_xL_z}}. \tag{8}$$

It can easily be verified that the $f_{n_x,n_z}(x,z)$ with integers $n_x \geq 1$ and $n_z \geq 1$ are orthonormal and satisfy the boundary conditions given by the hard-wall confinement. A wave function in three dimensions including the spin part can be expressed as a linear combination of $f_{n_x,n_z}(x,z)\exp\left(i\tilde{k}_y y\right)\zeta_{s,m_s}$, where $\tilde{k}_y$ is the wave number for the $y$ direction (along the nanowire). The $\zeta_{s,m_s}$ are spin states corresponding to the spin quantum number $s = 3/2$ and the quantum numbers $m_s \in \{\pm 3/2, \pm 1/2\}$ for the spin projection onto the $z$ axis.

In Fig. 5, the numerically obtained modulus squared (probability density) of a ground-state wave function $\psi(x,z)$ for a particle in a box of triangular cross section is compared with an analytical probability density. The numerical probability density is obtained by numerically solving the time-independent Schrödinger equation

$$\left[\frac{\hbar^2(k_x^2+k_z^2)}{2m}+V(x,z)\right]\psi(x,z) = E_g\psi(x,z), \tag{9}$$

where $E_g$ is the ground-state energy. We note that the result for $\psi(x,z)$ does not depend on the mass $m$ because of the hard-wall confinement. The chosen $V(x,z)$ is an infinite two-dimensional triangular potential well. More precisely, the obtuse isosceles triangle in Fig. 5 has a height of 3.8 nm and a base length of 38 nm, corresponding to the dimensions of the experimentally studied hut wire. The analytically obtained probability density is simply $|\psi(x,z)|^2 = |f_{1,1}(x,z)|^2$, see Eq. (8), using the dimensions $L_x = 11$ nm and $L_z = 3.6$ nm for a rectangular cross section. We compute the overlap of this analytical wave function $f_{1,1}(x,z)$ with the numerical wave function $\psi(x,z)$ for the above-mentioned dimensions. We find a relatively large overlap of 97.5% and thus assume that our way of modeling a particle in a triangular box with a particle in a rectangular box is valid as long as one tolerates deviations of a few percent (for more details see Sec. VII).

## IV. DIRECT RASHBA SPIN-ORBIT INTERACTION IN Ge HUT WIRES

Germanium and silicon are materials used to mass-produce electronic components. The fact that the unit cell of these materials has a center of inversion symmetry makes bulk Dresselhaus spin-orbit interaction zero in nanostructures consisting of stacked unit cells (however, Dresselhaus spin-orbit interaction can still exist due to the presence of material interfaces, for more information see Sec. VII). Furthermore, Rashba spin-orbit interaction in these materials is relatively weak. This is why it was long thought that Ge and Si are poor candidates for spintronics applications. However a recent study proved [11, 12] that holes in nanowires, particularly in Ge and Si, exhibit a Rashba type spin-orbit interaction unique to these systems. This direct Rashba spin-orbit interaction (DRSOI) is relatively strong already at weak and moderate electric fields, compared to the standard Rashba spin-orbit interaction, and can be $\sim 10$ meV in magnitude [11, 12].

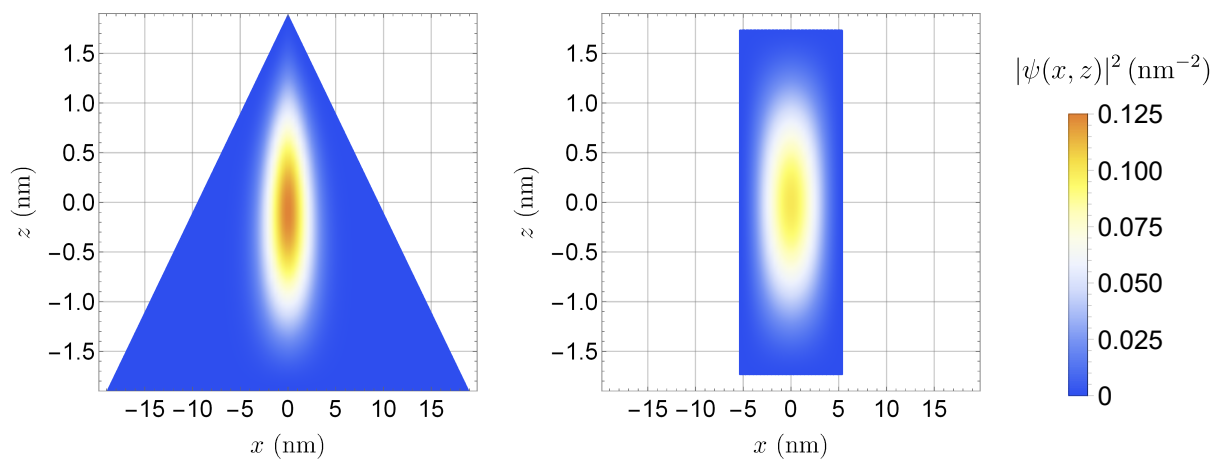

FIG. 5. The ground-state probability density for a triangular and rectangular cross section, respectively. Both plots have an aspect ratio of approximately 1:10. In order to model a hut wire whose cross section is close to an obtuse isosceles triangle, given a base length of 38 nm and a base/height aspect ratio of 10:1, we use $L_x = 11.0$ nm and $L_z = 3.6$ nm as the side lengths of a nanowire of rectangular cross section. As a consistency check we have numerically calculated the overlap between the ground-state wave functions of the obtuse isosceles triangle and the rectangle, resulting in a strong overlap of 0.975.

In the main body of the paper, we present experimental data for holes in a hut wire quantum dot. Among other things, our data show a pronounced $ST_-$ anticrossing. An avoided crossing between the singlet $S$ and the triplet $T_-$ can be caused by spin-orbit interaction [13]. More specifically, in the remainder of this Supplementary material, we determine that the $ST_-$ anticrossing is partly caused by the recently discovered "direct Rashba spin-orbit interaction" [11, 12].

The standard Rashba spin-orbit coupling for conduction band electrons is proportional to $(\boldsymbol{k} \times \boldsymbol{E}) \cdot \boldsymbol{\sigma}$, where $\boldsymbol{E} = E_x \hat{x} + E_y \hat{y} + E_z \hat{z}$ is the electric field, $\boldsymbol{\sigma} = \sigma_x \hat{x} + \sigma_y \hat{y} + \sigma_z \hat{z}$ is the vector of Pauli matrices for the electron spin, and $\hat{x}$, $\hat{y}$, $\hat{z}$ are the unit vectors for the $x$, $y$, $z$ directions [7]. Given a nanowire or an elongated quantum dot therein, a reasonable assumption is that the terms obtained with $\boldsymbol{k} = k_y \hat{y}$ have the strongest effects on the low-energy states. The unit vector $\hat{y}$ points along the nanowire and $k_y$ is the component of $\boldsymbol{k}$ along the wire. Consequently, the effective Rashba spin-orbit coupling is proportional to $(E_z \sigma_x - E_x \sigma_z) k_y$. If a magnetic field $\boldsymbol{B}$ is applied in the $z$ direction (see our experiment), the Zeeman term is proportional to $|\boldsymbol{B}| \sigma_z$. The $z$ axis therefore corresponds to the spin quantization axis: the triplet states $T_\pm$ have spin projections $m_s = \pm 1$ regarding the $z$ axis, whereas the singlet state $S$ has a spin projection $m_s = 0$. Since both the Zeeman term and $E_x \sigma_z k_y$ are proportional to $\sigma_z$, the spins of $T_\pm$ cannot be flipped by $E_x \sigma_z k_y$, and so $S$-$T_\pm$ crossings (instead of anticrossings) are expected for the Rashba spin-orbit interaction due to $E_x$. In contrast, the term $E_z \sigma_x k_y$, which is induced by the electric field component parallel to $\boldsymbol{B}$, can flip the spins and, hence, cause an avoided crossing between $S$ and $T_\pm$. In the following, considering the analogy to our experiment, we therefore study the direct Rashba spin-orbit interaction due to an electric field along $z$.

We start with the Hamiltonian

$$H = H_{\text{LK}}^{\text{Ge}} + H_{\text{BP}} + V - eE_z z. \tag{10}$$

The terms $H_{\text{LK}}^{\text{Ge}}$, $H_{\text{BP}}$, and $V$ are given in Eqs. (5), (6), and (7), respectively. The elementary positive charge is denoted by $e$. In order to estimate the strength of the direct Rashba spin-orbit interaction in the subband of lowest energy, we consider the eight basis states $|1, 1, 3/2\rangle$, $|1, 1, 1/2\rangle$, $|1, 1, -1/2\rangle$, $|1, 1, -3/2\rangle$, $|1, 2, 3/2\rangle$, $|1, 2, 1/2\rangle$, $|1, 2, -1/2\rangle$, and $|1, 2, -3/2\rangle$. These basis states are of the form $|n_x, n_z, m_s\rangle$ and can be represented by $f_{n_x, n_z}(x, z) \zeta_{s, m_s}$, where $n_x$ and $n_z$ are the orbital quantum numbers in the $x$ and $z$ direction, respectively, and $m_s$ with $s = 3/2$ is the spin-projection quantum number along the axis of strongest confinement, which in our case is the $z$ axis. The states with $m_s = \pm 1/2$ ($m_s = \pm 3/2$) are usually referred to as light-hole (heavy-hole) states. With our eight-dimensional subspace, we focus on the orbital ground state ($n_x = n_z = 1$) and the first excited orbital state that is coupled to the ground state via the electric field along $z$ ($n_x = 1$, $n_z = 2$) [11, 14]. We note that

$$\langle 1, 1, m_s | eE_z z | 1, 2, m_s \rangle = -\frac{16eE_z L_z}{9\pi^2}. \tag{11}$$

For the studied nanowire, the flatness of the rectangular cross section and the strain [see the Bir-Pikus Hamiltonian, Eq. (6), and the calculated strain distribution for the hut wire, Fig. 4] shift states with $m_s = \pm 3/2$ lower in energy compared to $m_s = \pm 1/2$. Thus, it is expected that the ground states in the hut wire will have a spin projection of approximately $\pm 3/2$ along the quantization axis $z$ [8] (see also Sec. VII).

By projecting the Hamiltonian of Eq. (10) onto the eight-dimensional subspace and applying a third-order Schrieffer-Wolff transformation, we perturbatively decouple the high-energy subspace $\{|1, 1, \pm 1/2\rangle, |1, 2, \pm 3/2\rangle, |1, 2, \pm 1/2\rangle\}$ from the lowest-energy states $|1, 1, \pm 3/2\rangle$, giving rise to the spin-orbit Hamiltonian

$$H_{\text{SO}}^{\text{DR}} = \alpha_{\text{DR}} E_z k_y \sigma_x, \tag{12}$$

which we derived assuming small $k_y$. The Pauli operator $\sigma_x$ acts on the two lowest-energy hole states. The prefactor $\alpha_{\text{DR}}$ is the effective Rashba coefficient for the direct Rashba spin-orbit interaction and given by the formula

$$\alpha_{\text{DR}} = \frac{2^{10} e \gamma_s^3 \hbar^4 L_x^2 L_z^4}{9(2\gamma_s - \gamma_1) \left[ 2\theta + 3\gamma_1 \pi^2 \hbar^2 L_x^2 + 2\gamma_s \pi^2 \hbar^2 (5L_x^2 - L_z^2) \right] \left[ \theta + \gamma_s \pi^2 \hbar^2 (2L_x^2 - L_z^2) \right]} \tag{13}$$

with

$$\theta = 2bL_x^2 L_z^2 m(\epsilon_{xx} - \epsilon_{zz}), \tag{14}$$

where $b$ is the deformation potential, $\epsilon_{xx}$ and $\epsilon_{zz}$ are the strain tensor elements, and $m$ is the free electron mass. Besides direct Rashba spin-orbit interaction other Rashba-like and Dresselhaus-like mechanisms can exist in our system. In particular, as our formalism cannot account for interface spin-orbit effects from a microscopic perspective, we are going to use the partly phenomenological spin-orbit Hamiltonian

$$H_{\text{SO}} = [(\alpha_{\text{DR}} + \alpha)E_z + \beta]\,k_y\sigma_x \tag{15}$$

in the remainder of the Supplementary material. Here, $\alpha_{\text{DR}}$ is the direct Rashba coefficient given by Eq. (13), $\alpha$ is a phenomenological constant which accounts for all spin-orbit mechanisms proportional to the electric field, excluding the DRSOI, and $\beta$ corresponds to all spin-orbit mechanisms which do not depend on the electric field.

## V. TWO-PARTICLE QUANTUM DOTS IN NANOWIRES

When electric gates lead to an additional confinement along the nanowire axis ($y$), a quantum dot is formed. We assume here that the electric gates generate an approximately parabolic potential. Furthermore, we assume that this gate-induced harmonic confinement in the $y$ direction is weaker than the hard-wall confinement in the $x$ and $z$ directions, which is reasonable because the effective $L_x$ and $L_z$ for hut wires are very small (see also Fig. 5). Thus, we can focus here on the simple case where the system is always in the orbital ground state regarding the $x$ and $z$ directions. In this section, we will derive the eigenfunctions for two charged particles in a one-dimensional (1D) harmonic oscillator potential, taking into account the repulsive Coulomb interaction. Our calculations are based on an approach by M. Taut [15], which we adjust to a 1D problem.

The two-particle Schrödinger equation is given by

$$\left(-\frac{\hbar^2}{2m^*}\frac{d^2}{dy_1^2} - \frac{\hbar^2}{2m^*}\frac{d^2}{dy_2^2} + \frac{m^*\omega^2 y_1^2}{2} + \frac{m^*\omega^2 y_2^2}{2} + \frac{e^2}{4\pi\epsilon_0\epsilon_r|y_1 - y_2|}\right)\psi(y_1, y_2) = E\psi(y_1, y_2). \tag{16}$$

Here, $y_1$ and $y_2$ are the coordinates of the first and second particle, respectively, $m^*$ is the effective mass of a particle, $\omega$ is the circular frequency for the parabolic confinement potential, $\epsilon_0$ is the vacuum permittivity, and $\epsilon_r$ is the relative permittivity of the material. Furthermore, $E$ stands for the eigenenergy and $\psi(y_1, y_2)$ for the two-particle wave function. It should be noted that the spin quantum number is omitted here because Eq. (16) is spin-independent. However, as we are interested in the wave functions of a singlet and a triplet state, we will consider the spin part in our final formulas for singlet and triplet eigenfunctions.

We introduce $y = y_1 - y_2$ and $Y = (y_1 + y_2)/2$. This change of variables allows us to separate Eq. (16) into two independent equations, one for $\xi(y)$ and another one for $\phi(Y)$, with $\psi(y_1, y_2) = \xi(y)\phi(Y)$ as the full wave function and $E = E_y + E_Y$ as the total energy. The first equation reads

$$\left(-\frac{\hbar^2}{2m^*}\frac{d^2}{dY^2} + 2m^*\omega^2 Y^2\right)\phi(Y) = 2E_Y\phi(Y). \tag{17}$$

Equation (17) is a harmonic-oscillator-like equation with the eigenenergies

$$E_Y = \hbar\omega(n_Y + 1/2), \quad n_Y \in \{0, 1, 2, \ldots\}, \tag{18}$$

and the eigenfunctions

$$\phi_{n_Y}(Y) = \frac{1}{\sqrt{2^{n_Y} n_Y!}}\left(\frac{2}{\lambda_c^2 \pi}\right)^{1/4}\exp\left(-\frac{Y^2}{\lambda_c^2}\right)H_{n_Y}\left(\sqrt{2}\frac{Y}{\lambda_c}\right), \tag{19}$$

where $H_{n_Y}(x)$ denotes the Hermite polynomial of order $n_Y$. The wave function for $n_Y = 0$ simplifies to

$$\phi_0(Y) = \left(\frac{2}{\lambda_c^2 \pi}\right)^{1/4}\exp\left(-\frac{Y^2}{\lambda_c^2}\right), \tag{20}$$

where $\lambda_c = \sqrt{\hbar/(m^*\omega)}$ is the confinement length of the quantum dot. The Schrödinger-type equation for $E_y$ and $\xi(y)$ is

$$\left(-\frac{\hbar^2}{2m^*}\frac{d^2}{dy^2} + \frac{m^*\omega^2 y^2}{8} + \frac{e^2}{8\pi\epsilon_0\epsilon_r|y|}\right)\xi(y) = \frac{E_y}{2}\xi(y). \tag{21}$$

Equation (21) is quasi-analytically solvable. The exact solution exists for special values of the confining potential with respect to the Coulomb repulsion terms, while approximate solutions exist in the limit of weak Coulomb repulsion (the so-called separated limit) and strong Coulomb repulsion (the so-called semi-classical limit). We adopt the latter approximation as the addition energy of our hut wire quantum dot is $\sim 2$ meV and is larger than the $ST_0$ energy splitting $\Delta E_{ST} \sim 200$ $\mu$eV, indicating that the Coulomb interaction is the largest energy scale in our system. The addition energy is estimated from Fig. 4b of the main text and is 8 mV, which translates to 2 meV when multiplied with a lever arm of $0.25e$, where $e$ is the elementary positive charge, and the $ST_0$ splitting is estimated from Fig. 4h of the main text.

In what follows we will expand the effective potential

$$V_{\text{eff}}(y) = \frac{m^* \omega^2 y^2}{8} + \frac{e^2}{8 \pi \epsilon_0 \epsilon_r |y|} \tag{22}$$

from Eq. (21) in a Taylor series around a point

$$y_0 = \pm \left( \frac{e^2}{2 \pi \epsilon_0 \epsilon_r m^* \omega^2} \right)^{1/3} \tag{23}$$

in which the first derivative of the effective potential vanishes. We note that $V_{\text{eff}}(y) = V_{\text{eff}}(-y)$. The plus sign in Eq. (23) is chosen when one wants to approximate $V_{\text{eff}}(y)$ near its minimum at positive $y$ (i.e., $y_1 > y_2$), whereas the minus sign is chosen for approximating $V_{\text{eff}}(y)$ near its minimum at negative $y$ (i.e., $y_1 < y_2$). The Taylor expansion around $y = y_0$ gives rise to the effective potential

$$\widetilde{V}_{\text{eff}}(y) = V_{\text{eff}}(y_0) + \frac{1}{2} \left. \frac{d^2 V_{\text{eff}}(y)}{dy^2} \right|_{y=y_0} (y - y_0)^2 \tag{24}$$

with

$$V_{\text{eff}}(y_0) = \frac{3}{8} \left( \frac{e^4 m^* \omega^2}{(2 \pi \epsilon_0 \epsilon_r)^2} \right)^{1/3} \tag{25}$$

and

$$\left. \frac{d^2 V_{\text{eff}}(y)}{dy^2} \right|_{y=y_0} = \frac{3 m^* \omega^2}{4}. \tag{26}$$

By replacing $V_{\text{eff}}(y)$ in Eq. (21) with $\widetilde{V}_{\text{eff}}(y)$, which is a very good approximation when $y \approx y_0$, i.e., when $y$ is near the considered minimum of $V_{\text{eff}}(y)$, we obtain the harmonic-oscillator-like equation

$$\left( -\frac{\hbar^2}{2 m^*} \frac{d^2}{dy^2} + \frac{3 m^* \omega^2}{8} (y - y_0)^2 \right) \xi(y) = \left( \frac{E_y}{2} - V_{\text{eff}}(y_0) \right) \xi(y), \tag{27}$$

resulting in the eigenenergies

$$E_y = \hbar \omega \sqrt{3} \left( n_y + \frac{1}{2} \right) + \frac{3}{4} \left( \frac{e^4 m^* \omega^2}{(2 \pi \epsilon_0 \epsilon_r)^2} \right)^{1/3} \tag{28}$$

with $n_y \in \{0, 1, 2, \ldots\}$. The eigenfunctions are

$$\xi_{n_y}(y) = \frac{1}{\sqrt{2^{n_y} n_y!}} \left( \frac{\sqrt{3}}{2 \pi \lambda_c^2} \right)^{1/4} \exp\left( -\frac{\sqrt{3}(y - y_0)^2}{4 \lambda_c^2} \right) H_{n_y} \left( \frac{3^{1/4}}{\sqrt{2}} \frac{y - y_0}{\lambda_c} \right), \tag{29}$$

where $H_{n_y}(x)$ denotes the Hermite polynomial of order $n_y$. The ground-state wave function corresponding to $n_y = 0$ is given by

$$\xi_0(y) = \left( \frac{\sqrt{3}}{2 \pi \lambda_c^2} \right)^{1/4} \exp\left( -\frac{\sqrt{3}(y - y_0)^2}{4 \lambda_c^2} \right). \tag{30}$$

The excited-state wave function corresponding to $n_y = 1$ is

$$\xi_1(y) = \frac{\sqrt{2}}{\pi^{1/4}} \left(\frac{\sqrt{3}}{2\lambda_c^2}\right)^{3/4} (y - y_0) \exp\left(-\frac{\sqrt{3}(y - y_0)^2}{4\lambda_c^2}\right). \tag{31}$$

It should be noted that all functions $\phi_{n_Y}(Y)$, given in Eq. (19), are symmetric under particle exchange because the coordinate $Y$ is symmetric under particle exchange. In contrast, $\xi_{n_y}(y)$ does not have symmetry when the particles are exchanged. So the symmetry of the total fermionic wave function will have to be set by choosing appropriate linear combinations of $\xi_{n_y}(y)$ and $\xi_{n_y}(-y)$. The totally antisymmetric singlet and triplet wave functions are thus given by

$$\langle Y, y \,|\, S, n_Y, n_y \rangle = \frac{1}{\sqrt{2}} \phi_{n_Y}(Y) \left[\xi_{n_y}(y) + \xi_{n_y}(-y)\right] \frac{|{\uparrow_1}{\downarrow_2}\rangle - |{\downarrow_1}{\uparrow_2}\rangle}{\sqrt{2}}, \tag{32}$$

$$\langle Y, y \,|\, T_{\pm,0}, n_Y, n_y \rangle = \frac{1}{\sqrt{2}} \phi_{n_Y}(Y) \left[\xi_{n_y}(y) - \xi_{n_y}(-y)\right] \begin{cases} |{\uparrow_1}{\uparrow_2}\rangle \\ \left(|{\uparrow_1}{\downarrow_2}\rangle + |{\downarrow_1}{\uparrow_2}\rangle\right)/\sqrt{2} \\ |{\downarrow_1}{\downarrow_2}\rangle. \end{cases} \tag{33}$$

Here, $\left(|{\uparrow_1}{\downarrow_2}\rangle - |{\downarrow_1}{\uparrow_2}\rangle\right)/\sqrt{2}$ is the spin part of the singlet wave function and is antisymmetric under particle exchange. Furthermore, $|{\downarrow_1}{\downarrow_2}\rangle$ is the spin part of the wave function of the $T_-$ state, $\left(|{\uparrow_1}{\downarrow_2}\rangle + |{\downarrow_1}{\uparrow_2}\rangle\right)/\sqrt{2}$ is the spin part of the wave function of the $T_0$ state and $|{\uparrow_1}{\uparrow_2}\rangle$ the spin part of the wave function of the $T_+$ state, all being symmetric with respect to the exchange of particles. In order to preserve the antisymmetric character of the total fermionic wave function, the spatial part of the wave function needs to be symmetric or antisymmetric under particle exchange and this is set by appropriately choosing a combination of $\xi_{n_y}(y)$, $\xi_{n_y}(-y)$ wave functions.

In the experiment, an anticrossing is measured between the singlet state $S$ and the $T_-$ triplet state. The ground-state singlet corresponds to the quantum numbers $n_Y = 0$ and $n_y = 0$. The energetically lowest $T_-$ state that gives rise to a $ST_-$ anticrossing is the triplet state described with the quantum numbers $n_Y = 0$ and $n_y = 1$.

Considering the results of this section and the regime $|y_0| \gg \lambda_c$, the normalized and antisymmetric wave functions of the ground-state singlet $|S, 0, 0\rangle$ and the energetically lowest triplet state which gives rise to a $ST_-$ anticrossing, $|T_-, 0, 1\rangle$, can be represented in position space by

$$\langle y_1, y_2 | S, 0, 0 \rangle = \frac{3^{1/8}}{\sqrt{2\pi}\lambda_c} e^{-\frac{(y_1 + y_2)^2}{4\lambda_c^2}} \left( e^{-\frac{\sqrt{3}(y_1 - y_2 - |y_0|)^2}{4\lambda_c^2}} + e^{-\frac{\sqrt{3}(y_2 - y_1 - |y_0|)^2}{4\lambda_c^2}} \right) \frac{|{\uparrow_1}{\downarrow_2}\rangle - |{\downarrow_1}{\uparrow_2}\rangle}{\sqrt{2}} \tag{34}$$

and

$$\langle y_1, y_2 | T_-, 0, 1 \rangle = \frac{3^{3/8}}{\sqrt{2\pi}\lambda_c^2} e^{-\frac{(y_1 + y_2)^2}{4\lambda_c^2}} \left( (y_1 - y_2 - |y_0|) e^{-\frac{\sqrt{3}(y_1 - y_2 - |y_0|)^2}{4\lambda_c^2}} - (y_2 - y_1 - |y_0|) e^{-\frac{\sqrt{3}(y_2 - y_1 - |y_0|)^2}{4\lambda_c^2}} \right) |{\downarrow_1}{\downarrow_2}\rangle. \tag{35}$$

We wish to emphasize again that the condition for the formulas derived in this and also the next section is that the Coulomb repulsion is the largest energy scale in the system, much larger than the single-particle level spacing, and so $|y_0|/\lambda_c \gg 1$. In the process of deriving and normalizing the wave functions in Eqs. (34) and (35), we have omitted terms which are suppressed by a factor of type $\exp\left(-y_0^2/\lambda_c^2\right)$. Consequently, the wave functions in Eqs. (34) and (35) are only approximately normalized to unity when $|y_0|/\lambda_c$ is finite.

## VI. THE $ST_-$ ANTICROSSING

In the absence of magnetic fields, the energy difference between the $T_-$ triplet state (with quantum numbers $n_Y = 0$ and $n_y = 1$, see Sec. V) and the singlet state (with $n_Y = n_y = 0$) is

$$\Delta E_{ST} = \sqrt{3}\hbar\omega, \tag{36}$$

as evident from Eq. (28). A magnetic field of strength $B = |\boldsymbol{B}|$ lowers the energy of $|T_-, 0, 1\rangle$ by the Zeeman energy $g\mu_B B$, where $g$ is the effective $g$-factor (considered as positive in our model) and $\mu_B$ the Bohr magneton. Thus, the energies of $|S, 0, 0\rangle$ and $|T_-, 0, 1\rangle$ are equal when $g\mu_B B = \Delta E_{ST}$. At this degeneracy point, the spin-orbit interaction results in a $ST_-$ anticrossing of magnitude

$$\Delta_{ST_-} = 2 \left| \langle S,0,0 | \left( H_{\text{SO}}^{(1)} + H_{\text{SO}}^{(2)} \right) | T_-, 0, 1 \rangle \right|. \tag{37}$$

We calculate $\Delta_{ST_-}$ by using the wave functions provided in Eqs. (34) and (35) for $|S,0,0\rangle$ and $|T_-,0,1\rangle$, respectively. For the spin-orbit interaction, we use the effective 1D term given in Eq. (15). More precisely, $H_{\text{SO}}^{(1)} + H_{\text{SO}}^{(2)}$ in Eq. (37) is obtained from $H_{\text{SO}}$ in Eq. (15) by replacing $k_y \sigma_x$ with $k_y^{(1)} \sigma_x^{(1)} + k_y^{(2)} \sigma_x^{(2)}$, where the superscripts (1) and (2) stand for the first and second particle, respectively.

Evaluating the integrals which occur in Eq. (37) is not straightforward. However, depending on the specific form of the integrand, these integrals can be separated into three types of integrals. They are then solvable with three different changes of variables, all corresponding to parametric equations of rotated ellipses:

$$
\text{(i)} \quad \begin{aligned}
y_1 &\to \rho \left( \sin(\varphi) + \frac{\cos(\varphi)}{3^{1/4}} \right) - \frac{|y_0|}{2}, \\
y_2 &\to \rho \left( \sin(\varphi) - \frac{\cos(\varphi)}{3^{1/4}} \right) + \frac{|y_0|}{2},
\end{aligned} \tag{38}
$$

$$
\text{(ii)} \quad \begin{aligned}
y_1 &\to \rho \left( \cos(\varphi) - \frac{\sin(\varphi)}{3^{1/4}} \right), \\
y_2 &\to \rho \left( \cos(\varphi) + \frac{\sin(\varphi)}{3^{1/4}} \right),
\end{aligned} \tag{39}
$$

$$
\text{(iii)} \quad \begin{aligned}
y_1 &\to \rho \left( \sin(\varphi) + \frac{\cos(\varphi)}{3^{1/4}} \right) + \frac{|y_0|}{2}, \\
y_2 &\to \rho \left( \sin(\varphi) - \frac{\cos(\varphi)}{3^{1/4}} \right) - \frac{|y_0|}{2}.
\end{aligned} \tag{40}
$$

In all three cases, the Jacobian is $2\rho/3^{1/4}$. The resulting formula for the $ST_-$ anticrossing is

$$\Delta_{ST_-} = 3^{1/4}\sqrt{2} \frac{|(\alpha_{\text{DR}} + \alpha)E_z + \beta|}{\lambda_c}. \tag{41}$$

We considered here the limit of strong Coulomb repulsion and assumed that electric-field-induced couplings are smaller than the orbital level spacing in the quantum dot. We recall that $E_z$ in Eq. (41) is the electric field along the axis of strongest confinement (out-of-plane direction, parallel to the applied magnetic field), $\alpha_{\text{DR}}$ is the direct Rashba coefficient derived in Sec. IV and given in Eq. (13), and $\lambda_c = \sqrt{\hbar/(m^*\omega)}$ is the confinement length of the quantum dot. We assumed that the quantum dot is elongated, i.e., the gate-induced confinement for the hut wire axis $y$ is weaker than the confinement for the transverse directions, which implies that $\lambda_c \gg L_x$ and $\lambda_c \gg L_z$.

An important quantity for many applications related to spin-orbit coupling is the spin-orbit length. Given an effective 1D Hamiltonian of type $\hbar^2 k_y^2/(2m^*) + \alpha' k_y \sigma_x$, where $\alpha'$ is a spin-orbit parameter, the spin-orbit length $\lambda_{\text{SO}}$ is usually defined as $\lambda_{\text{SO}} = \hbar^2/(m^*|\alpha'|)$ [12]. We therefore set $\alpha' = (\alpha_{\text{DR}} + \alpha)E_z + \beta$ and define

$$\lambda_{\text{SO}} = \frac{\hbar^2}{m^*|(\alpha_{\text{DR}} + \alpha)E_z + \beta|}. \tag{42}$$

Using Eq. (42) and the definition of $\lambda_c$, the formula in Eq. (41) for the magnitude $\Delta_{ST_-}$ of the $ST_-$ anticrossing can be written in many ways, such as

$$\Delta_{ST_-} = 3^{1/4}\sqrt{2} \frac{|\alpha'|}{\lambda_c} = \frac{3^{1/4}\sqrt{2}\hbar^2}{m^*\lambda_c\lambda_{\text{SO}}} = \frac{3^{1/4}\sqrt{2}\hbar\omega\lambda_c}{\lambda_{\text{SO}}} = \frac{\sqrt{2}\lambda_c g\mu_{\text{B}}B}{3^{1/4}\lambda_{\text{SO}}}. \tag{43}$$

We also used here our estimate that the $ST_-$ anticrossing occurs at $g\mu_{\text{B}}B = \Delta E_{ST} = \sqrt{3}\hbar\omega$, which is the point where the $|T_-,0,1\rangle$ and $|S,0,0\rangle$ introduced in Sec. V would cross in the absence of spin-orbit interaction. We note that the result $\Delta_{ST_-} \propto g\mu_{\text{B}}B\lambda_c/\lambda_{\text{SO}}$ seen in Eq. (43) is consistent with earlier studies of singlet-triplet anticrossings [13, 16].

Our theoretical estimation of the spin-orbit length is obtained with Eq. (42), which is independent of the confinement potential in the $y$ direction (along the hut wire). We use the formula

$$\lambda_{\text{SO}} = \frac{3^{1/4}\sqrt{2}\hbar\sqrt{\hbar\omega}}{\Delta_{ST_-}\sqrt{m^*}} \tag{44}$$

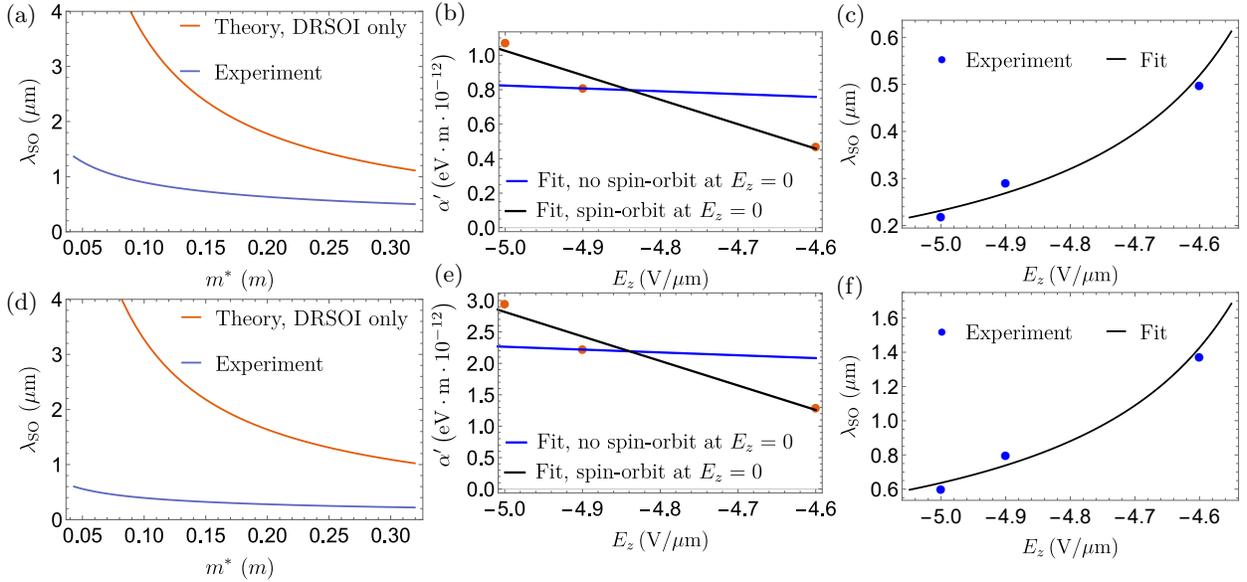

FIG. 6. (a, d) Spin-orbit length as a function of the effective mass, in the range of effective masses $0.042m \leq m^* \leq 0.32m$. Here, $L_x = 11$ nm, $L_y = 3.6$ nm, $\epsilon_{xx} = \epsilon_{yy} = -0.04$, $\epsilon_{zz} = 0.02$. The values of the $ST_0$ splitting $\Delta E_{ST} = 420, 242, 370$ $\mu$eV are read out from figures such as Fig. 5 (a) and (b) from the main body of the paper for the values of the electric field $E_z = -4.6, -4.9, -5.0$ V/$\mu$m respectively. The theory curve is a result calculated when only DRSOI is present in the system ($\alpha = \beta = 0$). Panel (a) is for $E_z = -4.6$ V/$\mu$m and panel (d) is for $E_z = -5$ V/$\mu$m. (b, e) The fitted spin-orbit parameter as a function of the electric field for the effective mass $m^* = 0.32m$ (b) and $m^* = 0.042m$ (e). (c, f) The fitted spin-orbit length $\lambda_{SO}$ as a function of the electric field for the effective mass $m^* = 0.32m$ (c) and $m^* = 0.042m$ (f).

for our experimental estimation of the spin-orbit length. As the effective mass is unknown in our system it is reasonable to assume that the light-hole mass $m_{LH} = m(\gamma_1 + 2\gamma_s)^{-1} \simeq 0.042m$ and the heavy-hole mass $m_{HH} = m(\gamma_1 - 2\gamma_s)^{-1} \simeq 0.32m$ of holes in bulk Ge are the lower and upper bounds for $m^*$ in our system. We note that an in-plane effective mass of approximately $m(\gamma_1 + \gamma_s)^{-1} \simeq 0.054m$ is expected for low-energy holes in 2D-like Ge nanostructures [17]. In Fig. 6 (a) and (d), we compare our theoretical estimation (for $\alpha = \beta = 0$) and experimental estimation of $\lambda_{SO}$ for the range $m_{LH} \leq m^* \leq m_{HH}$. We find that the theoretical and experimental results are different and that the quantitative agreement improves with increasing effective mass, indicating the importance of additional non-DRSOI spin-orbit mechanisms. To further characterize these mechanisms we fit Eq. (42) with $\alpha + \alpha_{DR}$ and $\beta$ being the free parameters to the measured data obtained by inserting the $ST_-$ anticrossing magnitude and orbital splitting (see Fig. 5 (a) and (b) in the main body of the paper) into Eq. (44) and display the results in Fig. 6 (b) and (e). For $m_{LH} \leq m^* \leq m_{HH}$ and $E_z = -5$ V/$\mu$m we obtain $2 \cdot 10^{-11}$ eV·m $\geq (\alpha_{DR} + \alpha)E_z \geq 7.1 \cdot 10^{-12}$ eV·m and $-1.7 \cdot 10^{-11}$ eV·m $\leq \beta \leq -6.1 \cdot 10^{-12}$ eV·m, with $|(\alpha_{DR} + \alpha)E_z/\beta| = 1.16$. Thus we conclude that the electric-field-dependent contribution $(\alpha_{DR} + \alpha)E_z$ is of a comparable order of magnitude to the electric-field-independent contribution $\beta$ but also of different sign, and $8\% \leq \alpha_{DR}E_z/[(\alpha + \alpha_{DR})E_z + \beta] \leq 22\%$, where $\alpha_{DR}$ is obtained from Eq. (13).

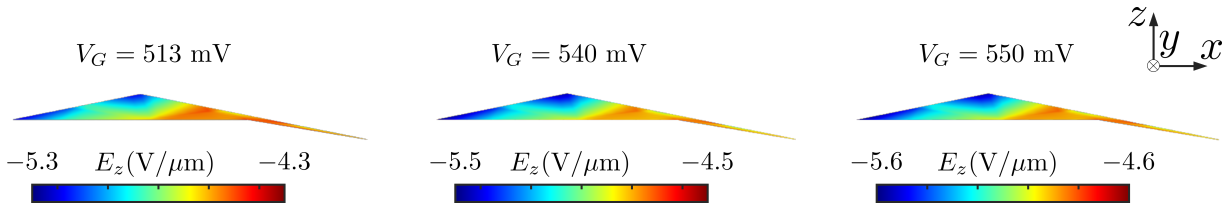

FIG. 7. COMSOL simulation of the electric field component $E_z$ along the $z$ direction for different gate voltages $V_G$. Because the electric field does not have a constant value throughout the wire we use the values $E_z = -4.6, -4.9, -5.0$ V/$\mu$m for $V_G = 513, 540, 550$ mV respectively.

The obtained spin-orbit length $200$ nm $\leq \lambda_{SO} \leq 600$ nm [see Fig. 6 (c) and (f) at $E_z = -5$ V/$\mu$m] is slightly larger than values reported for InAs nanowires (127 nm [13], 70 nm [18], 150 nm [19], 100–300 nm [20]), InSb nanowires (230 nm [21]), and cylindrical Ge/Si core/shell nanowires ($< 25$ nm [22]). There are two important reasons for the relatively large spin-orbit length in our experiment. First, the hut wire cross section, which resembles an obtuse isosceles triangle, leads to a hole confinement that is stronger in the $z$ than in the $x$ direction. Together with the strain in our sample, this results in a substantial heavy-hole–light-hole splitting. Consequently, as explained in Sec. II of Ref. [11], the direct Rashba spin-orbit interaction for the low-energy holes is expected to be weaker than, e.g., in the case of cylindrical symmetry [12]. The second reason is the effective electric field in the $z$ direction. The couplings that are proportional to $E_z L_z$ [e.g., Eq. (11)] are relatively small in our setup when $E_z \sim -5$ V/$\mu$m, which is the value we obtained from COMSOL simulations (Fig. 7). As evident from Eq. (42), the spin-orbit length depends strongly on $E_z$. We therefore expect that the spin-orbit interaction in our hut wires can be much enhanced with a modified setup where the strength of $E_z$ can be tuned over a wide range without significantly changing the gate-induced quantum dot confinement.

## VII. VALIDITY OF THE RESULTS

Our approach for calculating the direct Rashba coefficient $\alpha_{DR}$ relies on a Schrieffer-Wolff transformation. The Schrieffer-Wolff transformation can be performed if a low-energy subspace is sufficiently separated (decoupled) from states of higher energy. Similar to Ref. [8], our low-energy subspace consists of two heavy-hole states with spin $m_s = \pm 3/2$ along the quantization axis (out-of-plane direction). The strain in the Bir-Pikus Hamiltonian lifts the $m_s = \pm 1/2$ states much higher in energy compared to the $m_s = \pm 3/2$ states. For the four states of lowest energy, we find $1.49962 \leq |\langle J_z \rangle| \leq 1.49989$ for the absolute value of the spin expectation value along the quantization axis. This corresponds to a light-hole admixture of less than 1‰. For the estimation of $|\langle J_z \rangle|$, we considered the Hamiltonian $H_{LK}^{Ge} + H_{BP} + V$ [Eqs. (5) to (7)]. As a further consistency check we compared the low-energy spectrum of Eq. (12) with a numerically obtained spectrum of Eq. (10), including 3 orbital states in the $z$ direction and 3 orbital states in the $x$ direction. Both approaches gave rise to an approximately identical subband energy, validating the assumed low-energy subspace separation and small $k_y$ assumption we used to derive Eq. (12).

Replacing the triangular cross section with a rectangular cross section is an important simplification in our model. In order to find suitable values for the side lengths $L_x$ and $L_z$ of the rectangular cross section, we numerically calculate the ground-state wave function $\psi(x, z) = \chi_g(x, z)$ of a particle in an obtuse isosceles triangle with hard-wall confinement. The computed wave function is then compared with $\psi(x, z) = f_{1,1}(x, z)$ [Eq. (8)], which corresponds to the ground-state wave function of a particle in a rectangle with hard-wall confinement. Such a comparison is illustrated in Fig. 5. The overlap between $f_{1,1}(x, z)$ and $\chi_g(x, z)$ is obtained by calculating the integral $\langle f_{1,1} | \chi_g \rangle$. An overlap of absolute value 1 would mean that the two wave functions match perfectly. For $L_x = 11$ nm and $L_z = 3.6$ nm, the overlap is $\langle f_{1,1} | \chi_g \rangle = 0.975$ (see also Fig. 5), and so our approximation provides a good description of a particle in a triangular box up to a few percent.

In Secs. V and VI, we assume that the quantum dot is elongated ($\lambda_c \gg L_x$ and $\lambda_c \gg L_z$). Furthermore, we consider the limit of strong Coulomb repulsion ($|y_0| \gg \lambda_c$). Using $\hbar\omega \simeq 0.2$ meV (extracted from our experimental data) and $\epsilon_r \simeq 16$ for Ge, we obtain 93 nm $\geq \lambda_c \geq 34$ nm and 198 nm $\geq |y_0| \geq 102$ nm for the range $m_{LH} \leq m^* \leq m_{HH}$, which implies $2.1 \leq |y_0|/\lambda_c \leq 3$. In order to further test the validity of the approximation of strong Coulomb repulsion we have calculated the overlaps between the wave functions in Eq. (34) and Eq. (35) and the corresponding solutions obtained by numerically solving Eq. (16). We find that the lower bounds of the overlaps (corresponding to $m_{LH}$) are 0.976 for the ground state [Eq. (34)] and 0.82 for the excited one [Eq. (35)], justifying the applied approximation. It should be noted that one gets significantly higher overlaps when $m_{HH}$ is taken as the value of the effective mass. Moreover, given $L_x = 11$ nm and $L_z = 3.6$ nm as explained above, the assumption of an elongated quantum dot is justified.

The effective DRSOI term $H_{SO}^{DR} = \alpha_{DR} E_z k_y \sigma_x$ and its coefficient $\alpha_{DR}$ [Eq. (13)] for the lowest subband originate from the direct Rashba spin-orbit interaction and were derived by focusing on the regime of small $k_y$. From the fact that the calculated anticrossing $\Delta_{ST_-}$ and the measured value are of the same order of magnitude, we conclude that the spin-orbit coupling of holes in our hut wires is partly determined by the direct Rashba spin-orbit interaction. We note, however, that the dominance of the DRSOI over other spin-orbit mechanisms is expected to eventually decay when one changes from a system with two axes of strongest hole confinement (i.e., stronger heavy-hole–light-hole mixing) towards a system with only one axis of strongest hole confinement (i.e., weaker heavy-hole–light-hole mixing), as explained in Sec. II of Ref. [11]. Therefore, since $L_z$ is smaller than $L_x$ and $\lambda_c$ in our model, we are not surprised that other spin-orbit mechanisms are important. For example, when we add the standard Rashba term $r_{41}^{8v8v} E_z (J_y k_x - J_x k_y)$ [7] to the Hamiltonian $H$ of Eq. (10) and proceed as described in Sec. IV, we obtain an effective term $\alpha_R E_z k_y \sigma_x$ already in the second order of the Schrieffer-Wolff transformation. Using the bulk value $r_{41}^{8v8v} \simeq -0.4$ nm$^2 e$ [12], we find that $|\alpha_R/\alpha_{DR}|$ is almost 50% for our model parameters. That is, the direct Rashba term is significant for our hut wires, but is not as dominant as for previously studied Ge and Si nanowires [11, 12] whose cross sections are circles or squares, for instance. Additional contributions to the spin-orbit interaction can result from spin-orbit terms which are, e.g., cubic in the spin or the momentum [7, 23], from effective Dresselhaus terms due to the interfaces [7, 24–26], and from the fact that the hut wire cross-section does not have a center of inversion symmetry. The authors of Ref. [24] showed from a group theory approach that Si$_x$Ge$_{1-x}$/Si quantum wells can exhibit a Dresselhaus spin-orbit interaction due to the anisotropy of the chemical bonds at material interfaces. The authors of Refs. [25, 26] characterized this interface Dresselhaus spin-orbit interaction with tight-binding models. Including terms which are cubic in the spin or momentum, interface spin-orbit interaction or asymmetry of the cross-section goes beyond the scope of the formalism applied here and is planned as a forthcoming investigation.

## VIII. CALCULATION OF THE $ST_-$ ANTICROSSING AFTER UNITARY TRANSFORMATION

As a further consistency check of our results we try to obtain the $ST_-$ anticrossing with a method used in Ref. [13]. For our estimate of the $ST_-$ anticrossing, we consider the Hamiltonian

$$H = H_0^{(1)} + H_0^{(2)} + H_C + H_Z^{(1)} + H_Z^{(2)} + H_{SO}^{(1)} + H_{SO}^{(2)}, \tag{45}$$

where the superscripts in parentheses refer to the particle. We use

$$H_0 = \frac{\hbar^2 k_y^2}{2m^*} + \frac{m^* \omega^2 y^2}{2}, \tag{46}$$

which describes a particle in a parabolic potential, and the Zeeman term

$$H_Z = \frac{g\mu_B B}{2} \sigma_z, \tag{47}$$

with $g$ being the $g$-factor and $B$ the external magnetic field. The spin-orbit term

$$H_{SO} = \alpha' k_y \sigma_x \tag{48}$$

is the same as in Eq. (15). The Coulomb interaction is given by

$$H_C = \frac{e^2}{4\pi\epsilon_0\epsilon_r |y_1 - y_2|}. \tag{49}$$

The eigenstates of the Hamiltonian $H_0^{(1)} + H_0^{(2)} + H_C + H_Z^{(1)} + H_Z^{(2)}$ relevant for our calculation are already given in Eqs. (34) and (35). In difference to our approach, the authors in Refs. [13] and [16] rely on the Hund-Mulliken method of atomic orbitals and on variational calculus, respectively, in order to calculate the eigenstates.

Following earlier work [13, 16, 27–29], the Hamiltonian in Eq. (45) can be transformed with a unitary operator

$$U = \mathbb{1} - i\frac{m^*\alpha'}{\hbar^2}\left(y_1\sigma_x^{(1)} + y_2\sigma_x^{(2)}\right) + \cdots, \tag{50}$$

leading to

$$\widetilde{H} = U^\dagger H U = H_0^{(1)} + H_0^{(2)} + H_C + H_Z^{(1)} + H_Z^{(2)} + H_{SO,Z}^{(1)} + H_{SO,Z}^{(2)} + \cdots, \tag{51}$$

where

$$H_{SO,Z} = \frac{m^*\alpha' g\mu_B B}{\hbar^2} y\sigma_y. \tag{52}$$

The unitary transformation described by the operator $U$ removes the term $H_{SO}$ and generates a combined spin-orbit and Zeeman term $H_{SO,Z}$, depending on the position operator, making our calculations much more straightforward. The transformation is particularly useful when $\lambda_{SO} \gg |y_0|$, $\lambda_{SO} \gg \lambda_c$, and $g\mu_B B \lesssim \hbar\omega$. Given these assumptions, additional terms in Eqs. (50) and (51) are relatively small and have been replaced by "$\cdots$".

We can now calculate the magnitude of the $ST_-$ anticrossing as follows,

$$\Delta_{ST_-} = 2\left|\langle S,0,0|\left(H_{SO,Z}^{(1)} + H_{SO,Z}^{(2)}\right)|T_-,0,1\rangle\right|, \tag{53}$$

with $|S, 0, 0\rangle$ and $|T, 0, 1\rangle$ being given by Eq. (34) and Eq. (35), respectively. We obtain

$$\Delta_{ST_-} = 2 \frac{g\mu_B B}{\sqrt{2}\lambda_{SO}} r_{12}.$$ (54)

This is the same result as in Ref. [13] up to a factor of 2 occurring from a different definition of the anticrossing as compared to our study. The introduced quantity

$$r_{12} = \left| \int_{-\infty}^{\infty} \int_{-\infty}^{\infty} dy_1 dy_2 \left[ \psi^T(y_1, y_2) \right]^* (y_1 - y_2) \psi^S(y_1, y_2) \right|$$ (55)

is a coupling parameter which has the units of distance. In Eq. (55), the $\psi^S(y_1, y_2)$ and $\psi^T(y_1, y_2)$ are the orbital parts of the singlet and triplet wave functions explicitly given in Eq. (34) and Eq. (35). The authors of Refs. [13, 16] report $r_{12} \approx \lambda_c$. In our case, the result in the limit $|y_0| \gg \lambda_c$ is

$$r_{12} = \frac{\lambda_c}{3^{1/4}},$$ (56)

which is in accordance with previous studies [13, 16], up to a numerical prefactor. Finally, the anticrossing is

$$\Delta_{ST_-} = \sqrt{2} \frac{g\mu_B B}{\lambda_{SO}} \frac{\lambda_c}{3^{1/4}},$$ (57)

which is the same result as in Eq. (43) and in agreement with the results reported in Refs. [13, 16].

---